\documentclass[11.5pt]{article}

\usepackage[utf8]{inputenc}
\usepackage[margin=1.0in]{geometry}

\usepackage{upgreek}
\usepackage{graphicx}
\usepackage{float}
\graphicspath{ {Images/} }

\usepackage{cite}
\usepackage{braket}
\usepackage{amsmath}
\usepackage{gensymb}
\usepackage{multicol}
\usepackage{authblk}
\usepackage{hyperref}

\newcommand{\beginsupplement}{%
        \setcounter{table}{0}
        \renewcommand{\thetable}{S\arabic{table}}%
        \setcounter{figure}{0}
        \renewcommand{\thefigure}{S\arabic{figure}}
        \setcounter{equation}{0}
        \renewcommand{\theequation}{S\arabic{equation}}%
     }

\title{Explicit models of motions to understand protein side-chain dynamics}
\author[1]{Nicolas Bolik-Coulon}
\author[2]{Olivier Languin-Catto{\"{e}}n}
\author[1]{Diego Carnevale}
\author[1,2]{Milan Zachrdla}
\author[3]{Damien Laage}
\author[2]{Fabio Sterpone}
\author[2]{Guillaume Stirnemann}
\author[1]{Fabien Ferrage}

\affil[1]{Laboratoire des Biomolécules, LBM, Département de chimie, École Normale Supérieure, PSL University, Sorbonne Université, CNRS, 24 rue Lhomond, 75005 Paris, France}
\affil[2]{Laboratoire de Biochimie Théorique, CNRS, Institut de Biologie Physico-Chimique, Sorbonne Paris Cité, PSL University, 13 rue Pierre et Marie Curie, 75005 Paris, France}
\affil[3]{PASTEUR, Département de chimie, École Normale Supérieure, PSL University, Sorbonne Université, CNRS, 24 rue Lhomond, 75005 Paris, France}

\begin{document}

\date{}

\maketitle


\begin{abstract}
Nuclear magnetic relaxation is widely used to probe protein dynamics. For decades, most analyses of relaxation in proteins have relied successfully on the model-free approach, forgoing mechanistic descriptions of motions. Model-free types of correlation functions cannot describe a large carbon-13 relaxation dataset in protein sidechains. Here, we use molecular dynamics simulations to design explicit models of motion and solve Fokker-Planck diffusion equations. These models of motion provide better agreement with relaxation data, mechanistic insight and a direct link to configuration entropy.
\end{abstract}

Protein dynamics can be studied in the liquid state using a wide range of spectroscopic and scattering approaches \cite{Roy_NatMet_2008,Charlier_ChemSocRev_2016,Grimaldo_QRB_2019,Tumbic_ARAC_2021}. For example, fluorescence anisotropy decay curves are direct measurements of the time-correlation function for the orientation of the fluorescence probe axis frame \cite{Kinosita_BiophysJ_1977}. Nuclear Magnetic Resonance (NMR) relaxation rates probe the spectral density function, the Fourier transform of the time-correlation function for the orientation of the axis frames of spin-interactions \cite{Wangsness_PhysRev_1953,Redfield_IBM_1957,Palmer_ChemRev_2004,Nicholas_PNMRS_2010} in the fixed laboratory frame. These methods are sensitive to both the overall rotational diffusion of the molecule and internal motions of small, well-defined moieties. \\
The quantitative interpretation of relaxation rates in terms of motion requires parametrized models of correlation functions. A diversity of such models of motions were introduced in the second half of the XX$^\mathrm{th}$ century: rotation on a cone \cite{Wallach_JCP_1967}, orientation jump \cite{Wittebort_JCP_1978} and diffusion in a cone \cite{Kinosita_BiophysJ_1977}. Discriminating between these models is difficult, particularly when limited experimental datasets are available, which motivated the introduction of simplified correlation functions. For instance, in NMR spectroscopy, the Model Free (MF) approach approximates the correlation function for internal motions $C_{int}(t)$ to a single exponential decay term \cite{Lipari_JACS_1982, Halle_JCP_2009}:
\begin{equation}
C_{int}(t) = \mathcal{S}^2 + (1 - \mathcal{S}^2)e^{-t/\tau},
\end{equation}
with $\mathcal{S}^2$ the generalized squared order parameter and $\tau$ an effective correlation time. The great simplification brought by the MF approach has led to hundreds of successful analyses of NMR relaxation data recorded on proteins \cite{Jarymowycz_CR_2006} and a better understanding of internal protein motions. \\
A variety of NMR methods and new instruments make it possible to record extensive sets of relaxation rates, and to probe the range of validity of the MF approach.  We have recently reported site-specific isoleucine-$\delta$1 carbon-13 relaxation measurements recorded on the protein Ubiquitin over two orders of magnitude of magnetic fields \cite{Cousin_JACS_2018,BolikCoulon_JMR_2020}. Correlation functions derived from the MF approach were unable to describe both auto- and cross-correlated relaxation rates \cite{BolikCoulon_JMR_2020}.  The MF approach was a necessary simplification four decades ago but, today, the nature of motions is better known thanks to Molecular Dynamics (MD) simulations. Would explicit models of motions provide a more accurate and informative framework to analyze extensive relaxation data sets? Here, we use  MD simulations to define key properties of motions (\textit{e.g.}, to identify which rotamer states are populated), and write Fokker-Planck diffusion equations that fulfill these properties. We employ the resulting correlation functions to analyze NMR relaxation data and to quantitatively describe the dynamics of protein side-chains and the underlying molecular mechanisms.\\

\begin{figure*}[!ht]
\begin{center}
\includegraphics[width=0.65\textwidth]{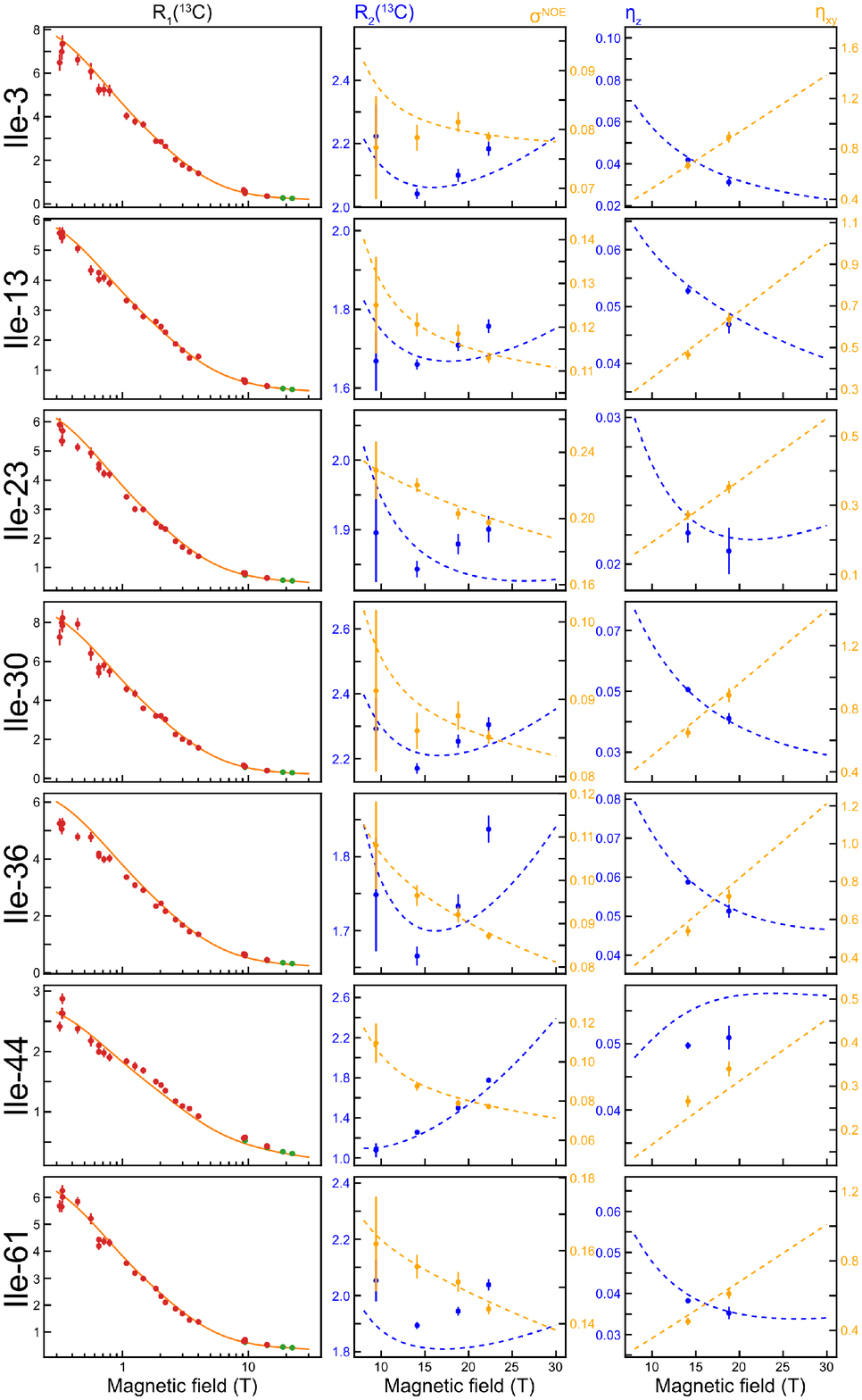}
\end{center}
\caption{\label{fig:MFfits}
NMR relaxation data analysis for the 7 Ile-$\delta$1 of Ubiquitin using a MF-type of correlation function. The R$_2$ of Ile-44 were included in the MCMC procedure and an exchange term was added to the R$_2$: $R_2' (B_0)= \alpha_{ex}B_0^2 + R_2(B_0)$ were $B_0$ is the magnetic field, $R_2$ contains contributions from the DD and CSA interactions, and $\alpha_{ex}$ accounts for the chemical exchange contribution and is a free parameter in the MCMC procedure: $\alpha_{ex} = 1.7 \pm 0.1$\,ms.T$^{-2}$. In the R$_1$ relaxation dispersion profiles, the high-field and corrected relaxometry data [20] are respectively shown in green and red.}
\end{figure*}

\begin{figure*}[!ht]
\begin{center}
\includegraphics[width=0.7\textwidth]{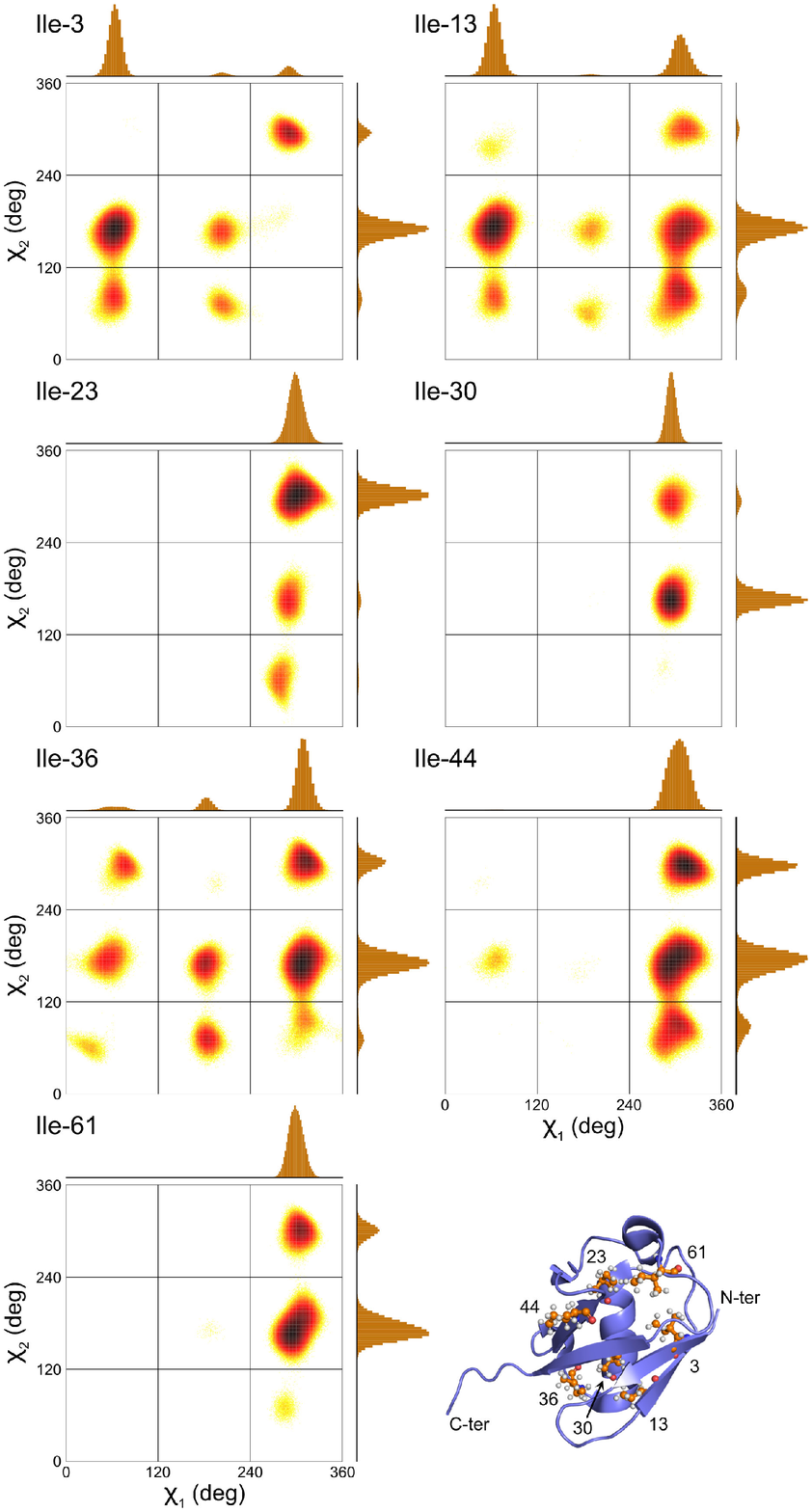}
\end{center}
\caption{\label{fig:MDdistributions}
Probability distributions for the dihedral angles $\chi_1$ and $\chi_2$ of the seven isoleucine side-chains determined from the MD trajectory and structure of Ubiquitine (PDB 1D3Z) showing the 7 isoleucines.}
\end{figure*}

\begin{figure*}[!ht]
\begin{center}
\includegraphics[width=0.75\textwidth]{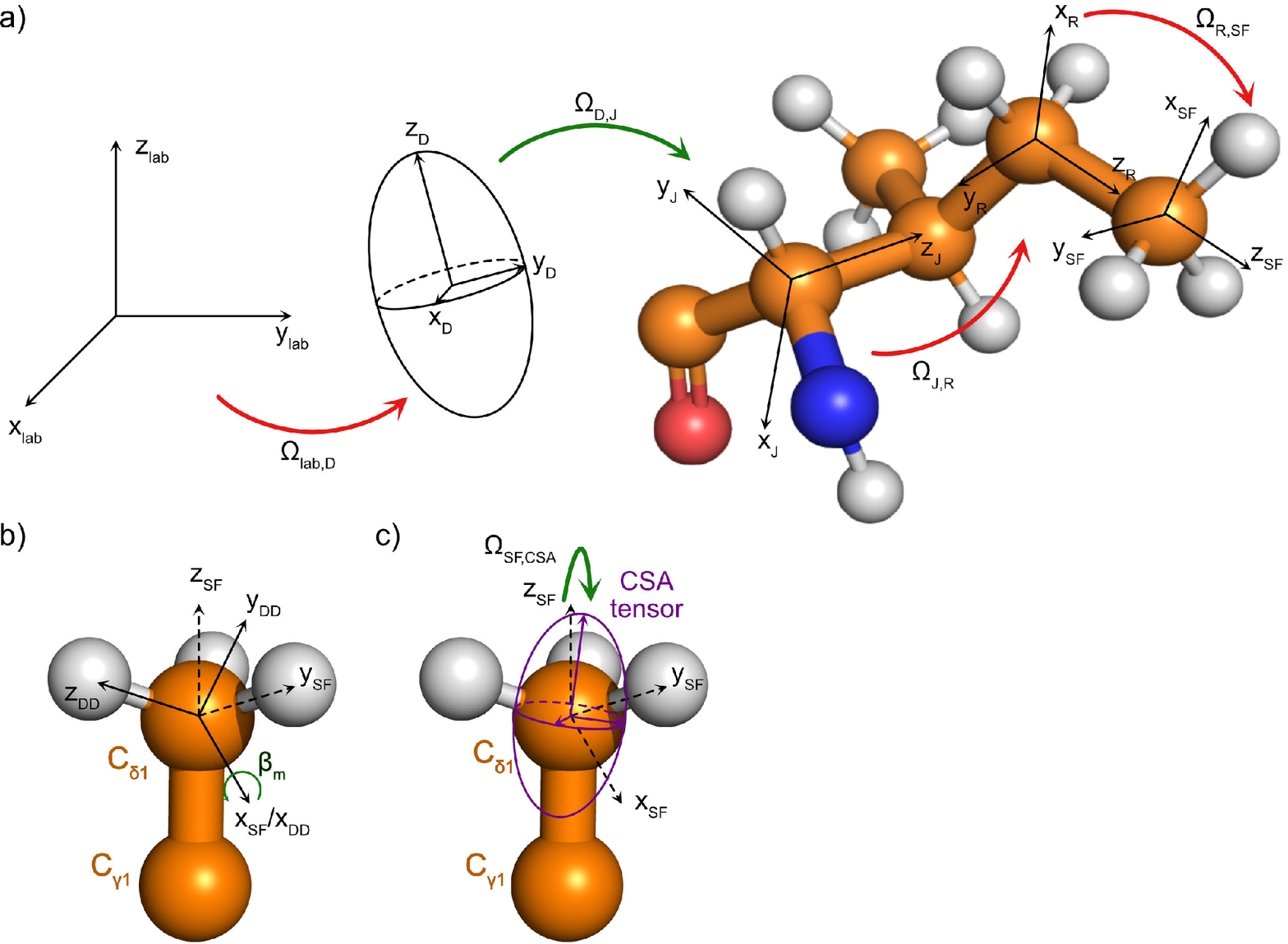}
\end{center}
\caption{\label{fig:FrameTransformation}
Frame transformation used to write the correlation function. \textbf{a)} The laboratory frame is a fixed frame. The diffusion frame (D) is associated to the diffusion tensor of the whole protein. The transformation from the laboratory to the diffusion frame is used to describe the overall tumbling. The jump frame (J) is a local frame for the amino acid of interest and is considered fixed in the diffusion frame. The rotamer frame (R) is defined for each possible rotamer. The transformation from the jump frame to the rotamer frame is used to describe rotamer jumps. In the case of correlation functions involving dipolar interactions, a system frame (SF) is defined with its main axis aligned along the C$_{\gamma1}$-C$_{\delta1}$ bond and the x-axis follows the direction of one C-H bond such that the transformation from the rotamer to system frame describes the methyl-group rotation. The interaction frames are fixed in the system frame (or in the rotamer frame for the CSA interactions) and are not shown in (b) for the dipole-dipole interaction and (c) for the CSA interaction. The time-dependent (time-independent) transformations are shown with red (green) arrows. The SF are shown with the dashed lines in (b) and (c). In the case of the carbon-proton/deuterium dipolar interactions, the x-axis of the SF and interaction frames are identical.}
\end{figure*}

\begin{figure*}[!ht]
\begin{center}
\includegraphics[width=0.7\textwidth]{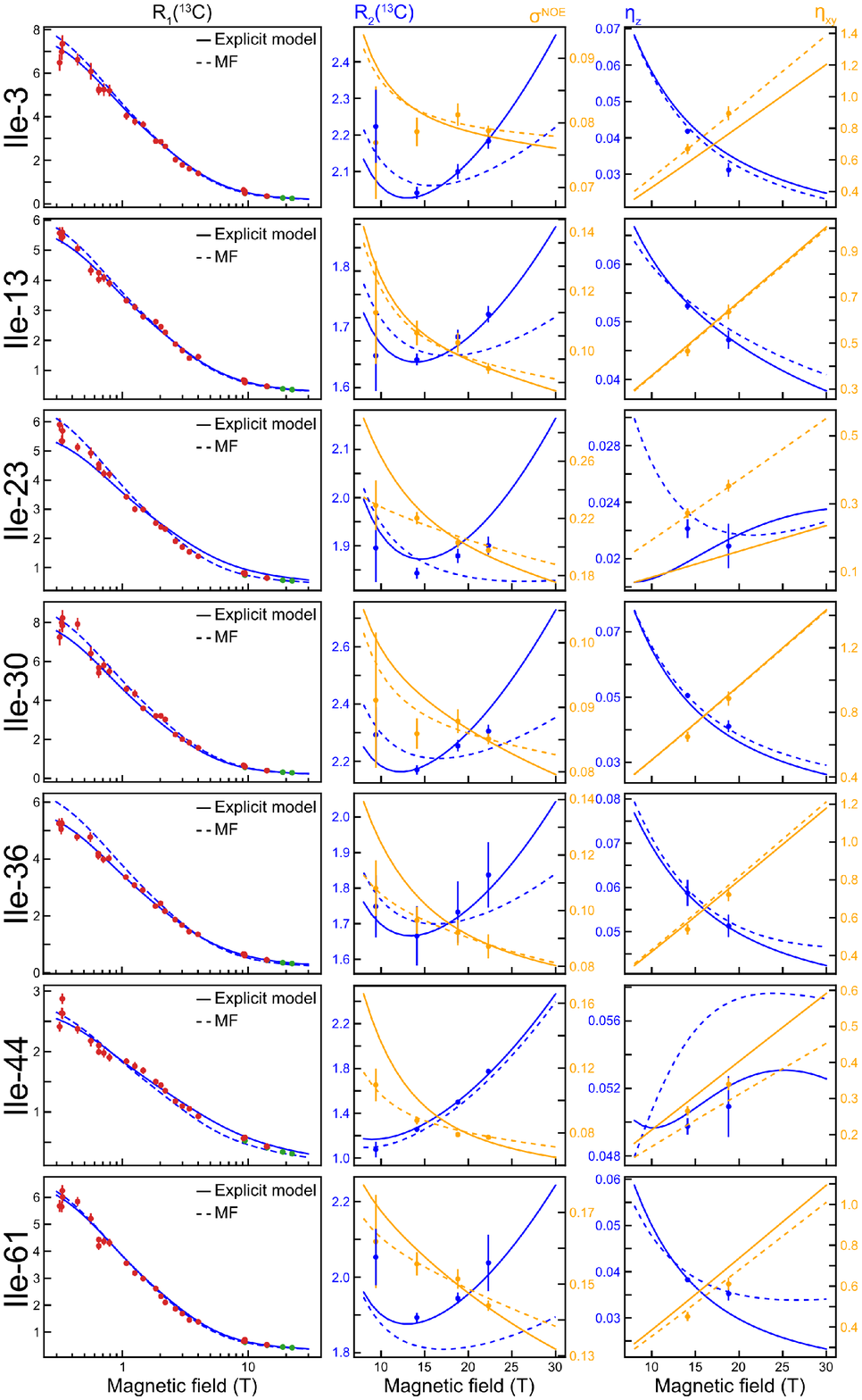}
\end{center}
\caption{\label{fig:FitsAll}
NMR relaxation data analysis for the 7 Ile-$\delta$1 of Ubiquitin using an explicit model of motion where the rotation of the methyl group is defined by a uniform diffusion constant for all rotamers of each side-chain (solid lines). The results obtained from the MF analysis are shown as dashed lines. The R$_2$ of Ile-44 were included in the MCMC procedure and an exchange term was added to the R$_2$: $R_2' (B_0)= \alpha_{ex}B_0^2 + R_2(B_0)$ were $B_0$ is the magnetic field, $R_2$ contains contributions from the DD and CSA interactions, and $\alpha_{ex}$ accounts for the chemical exchange contribution and is a free parameter in the MCMC procedure: $\alpha_{ex} = 1.3 \pm 0.2$\,ms.T$^{-2}$. In the R$_1$ relaxation dispersion profiles, the high-field and corrected relaxometry data are respectively shown in green and red.}
\end{figure*}

\begin{figure*}[!ht]
\begin{center}
\includegraphics[width=0.9\textwidth]{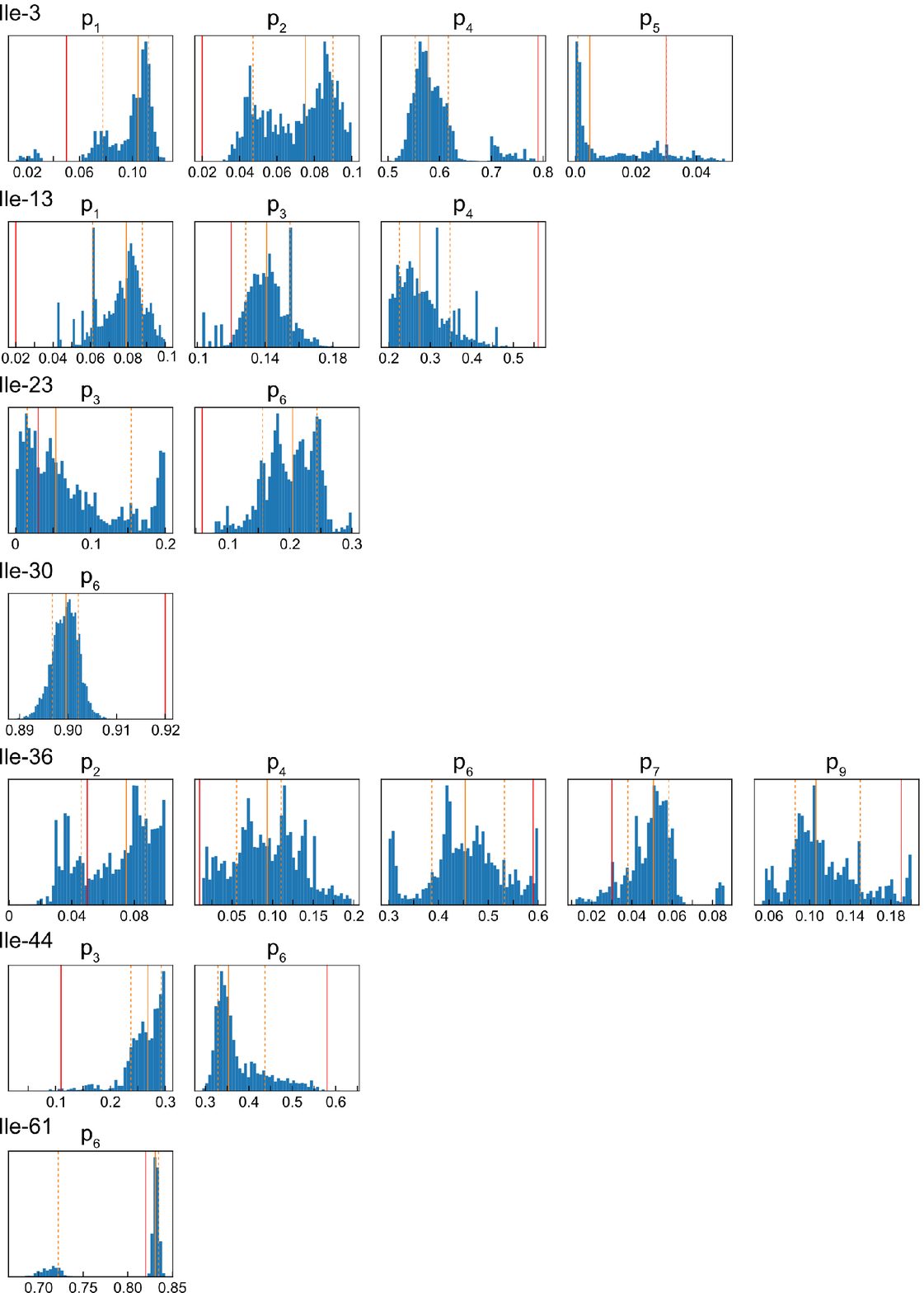}
\end{center}
\caption{\label{fig:PopDistributions}
Distributions of rotamer populations after the MCMC analysis of NMR relaxation using explicit models of motions. The values obtained from the MD simulation (Table\,\ref{table:PopMD}) are shown with the red solid line. Mean and width of the distributions are shown with the orange solid and dash vertical lines respectively.}
\end{figure*}

\begin{figure*}[!ht]
\begin{center}
\includegraphics[width=0.9\textwidth]{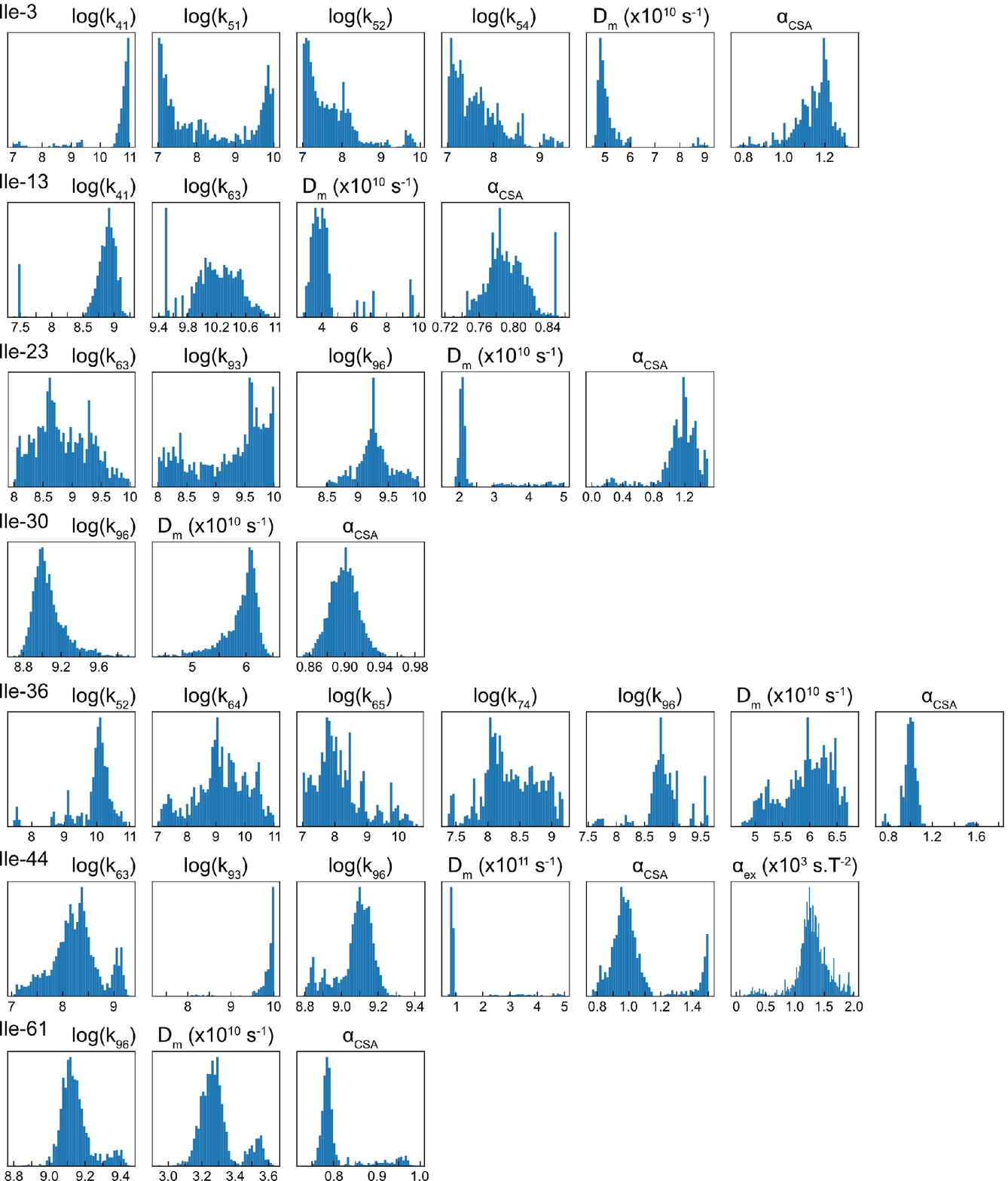}
\end{center}
\caption{\label{fig:DynDistributions}
Distributions of the decimal logarithm of jump rates (expressed in s$^{-1}$), diffusion coefficient for methyl rotation D$_\mathrm{rot}$, scaling factor for the CSA amplitudes and, in the case of Ile-44, the exchange contribution to relaxation $\alpha_{ex}$ after the MCMC analysis of NMR relaxation using explicit models of motions with identical values of $D_m$ in all rotamers.}
\end{figure*}

\begin{figure*}[!ht]
\begin{center}
\includegraphics[width=0.8\textwidth]{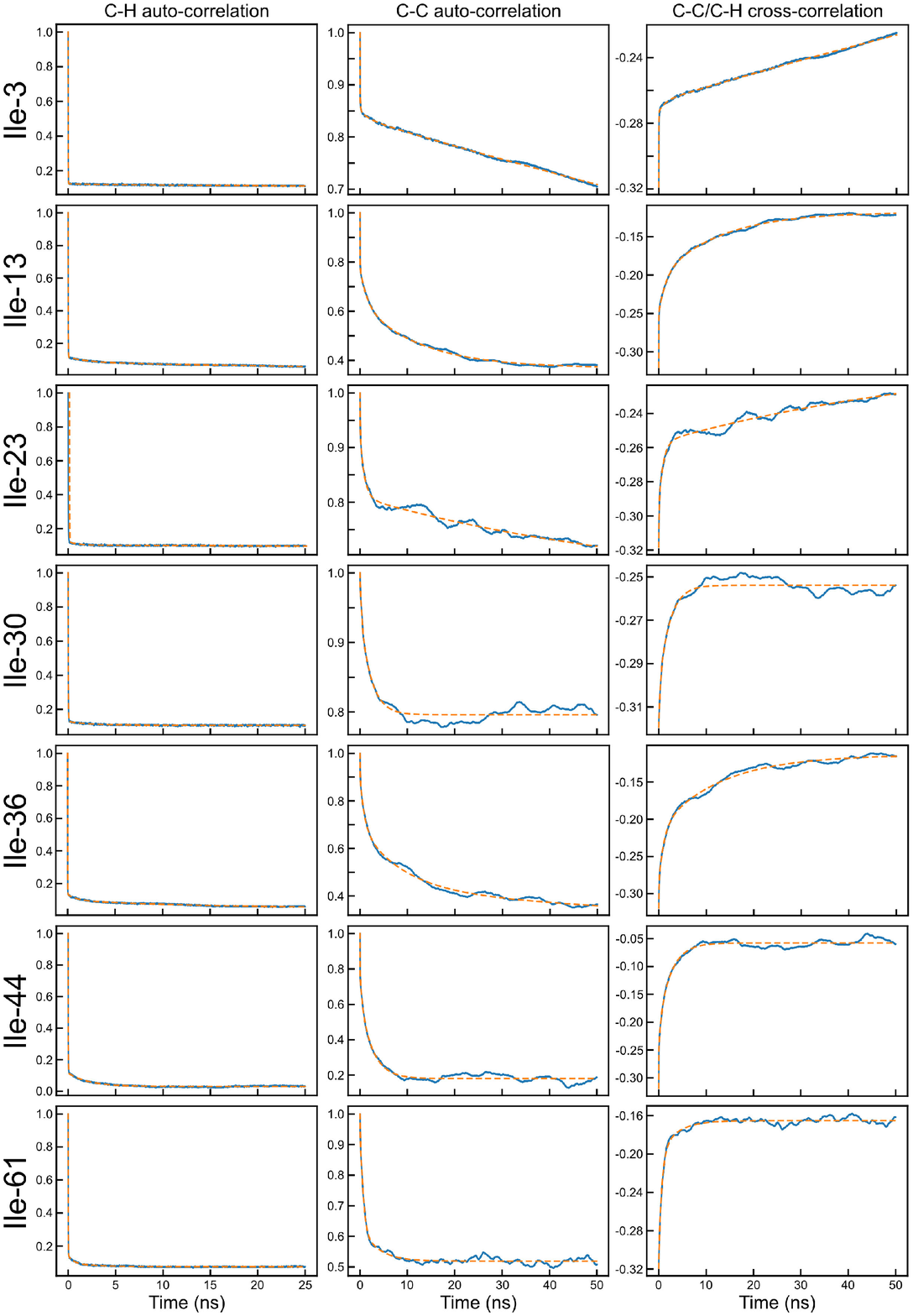}
\end{center}
\caption{\label{fig:MDcorrFuncFits}
Fits (orange, dash) of the MD correlelation function (blue, plain) for the C$_{\delta1}$-H auto-correlation, C$_\gamma$-C$_{\delta1}$ auto-correlation and C$_\gamma$-C$_{\delta1}$/ C$_{\delta1}$-H cross-correlation.}
\end{figure*}

\begin{figure*}[!ht]
\begin{center}
\includegraphics[width=0.7\textwidth]{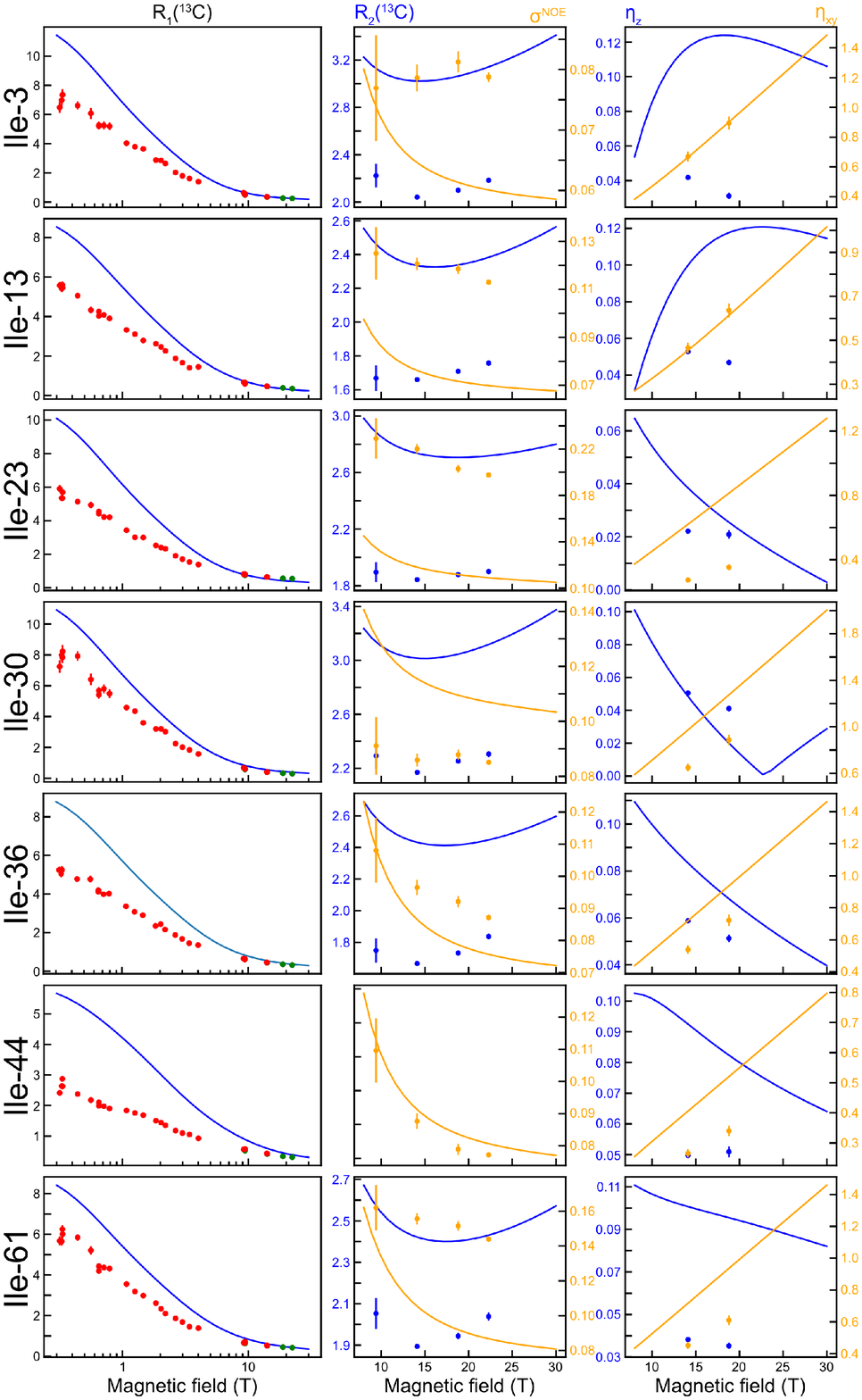}
\end{center}
\caption{\label{fig:MDrates_MF}
NMR relaxation rates calculated using the MD correlation functions. The MD correlation functions were fitted to a sum of 4 decaying exponential terms $C(t) = C_0 + \sum_{i=1}^4 e^{-t/\tau_i}$ and used to compute the relaxation rates. We use an isotropic overall diffusion tensor with global tumbling correlation time $\tau_c=5.028$\,ns. In the R$_1$ relaxation dispersion profiles, the high-field and corrected relaxometry data are respectively shown in green and red. For these calculations, the carbon-13 CSA was supposed to be symmetric and aligned along the C$_{\gamma1}$-C$_{\delta1}$ bond with value equal to the averaged CSA value weighted with the rotamer populations obtained from the MD simulation.}
\end{figure*}

\begin{figure*}[!ht]
\begin{center}
\includegraphics[width=0.7\textwidth]{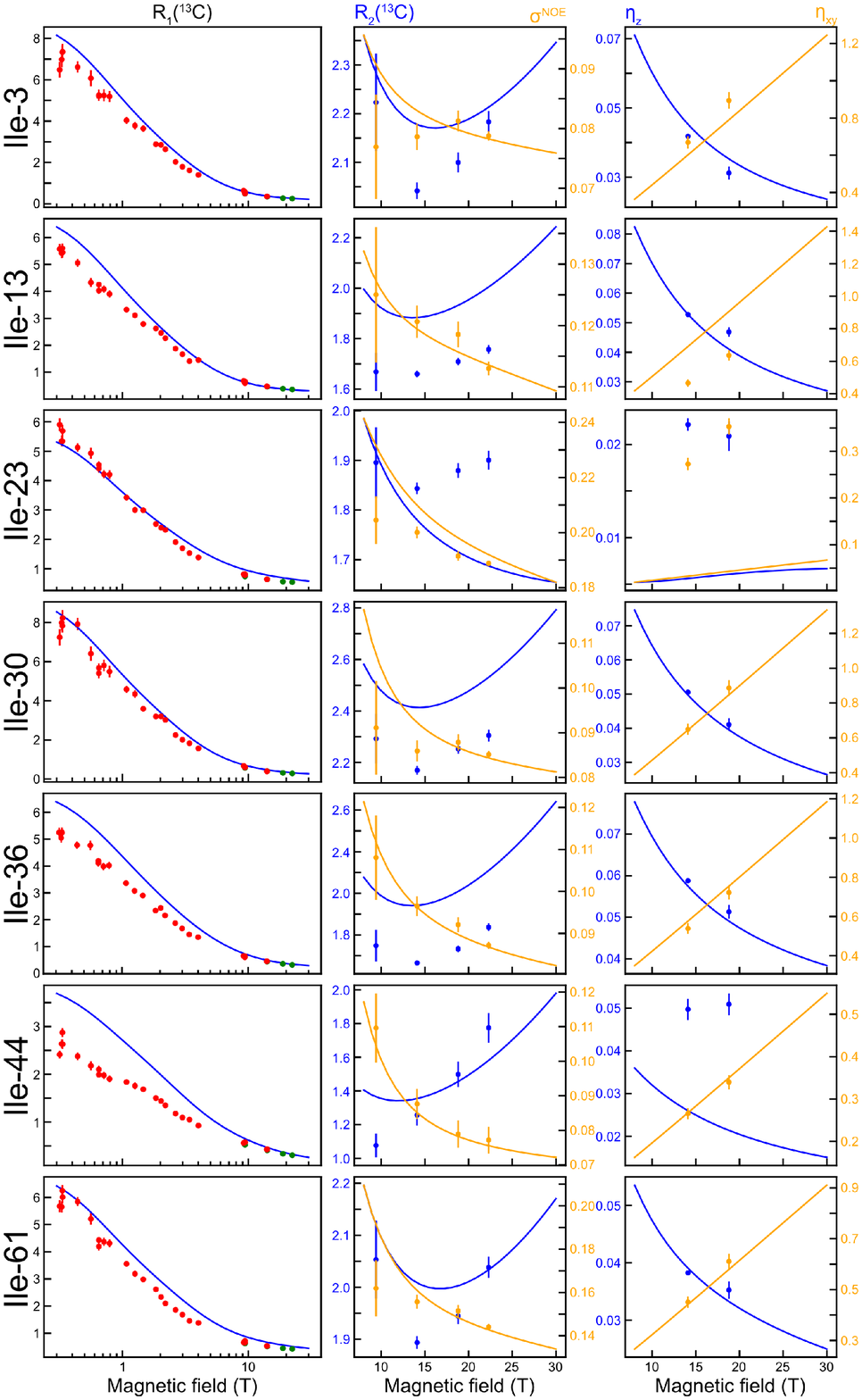}
\end{center}
\caption{\label{fig:MDrates_Explicit}
NMR relaxation rates calculated using information obtained from the MD simulations. We used an explicit model of correlation function with population and jump rates fixed to their MD values. Only the diffusion coefficient for methyl rotation and the scaling factor for the CSA amplitudes were determined using an MCMC procedure on carbon longitudinal R$_1$ and transverse R$_2$ relaxation rates, carbon-proton DD cross-relaxation rates $\sigma^\mathrm{NOE}$ as well as longitudinal and transverse CSA/DD cross-correlated cross-relaxation rates $\eta_z$ and $\eta_{xz}$. The overall diffusion tensor is axially symmetric, as reported in the main text from the analysis of experimental $^{15}$N-relaxation data. In the R$_1$ relaxation dispersion profiles, the high-field and corrected relaxometry data are respectively shown in green and red. The R$_2$ of Ile-44 were included in the MCMC procedure and an exchange term was added to the R$_2$: $R_2' (B_0)= \alpha_{ex}B_0^2 + R_2(B_0)$ were $B_0$ is the magnetic field, $R_2$ contains contributions from the DD and CSA interactions, and $\alpha_{ex}$ accounts for the chemical exchange contribution and is a free parameter in the MCMC procedure: $\alpha_{ex} = 0.7 \pm 0.3$\,ms.T$^{-2}$.}
\end{figure*}

\begin{figure*}[!ht]
\begin{center}
\includegraphics[width=0.9\textwidth]{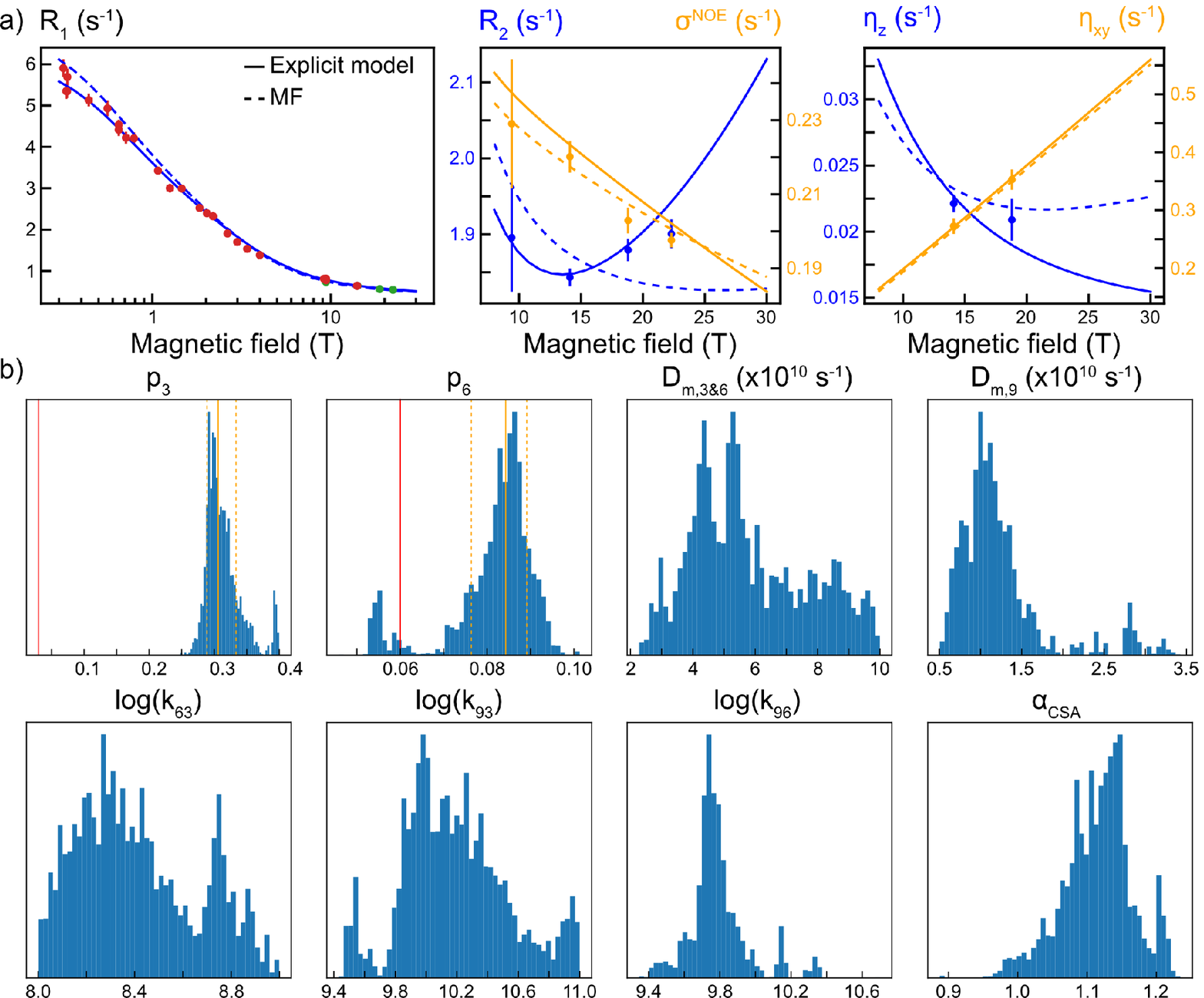}
\end{center}
\caption{\label{fig:Fit23}
Results of the analysis using explicit models of motions with methyl rotation correlated with the rotamer state for Ile-23. a) Measured and calculated relaxation rates following a MCMC procedure using the explicit model with correlated jumps and rotation (solid) or MF-type of correlation function (dash). The high-field R$_1$ and the corrected relaxometry rates are shown in green and red respectively. b) Distributions of the parameters of the explicit model of motions after the MCMC analysis: p$_i$ ais the population of rotamer $i$, k$_{ij}$ is the jump rate from rotamer $jj$ to rotamer $i$ (expressed in s$^{-1}$), D$_\mathrm{m,3\&6}$ is the diffusion coefficient for methyl-group rotation in rotamers 3 and 6 (supposed to be identical in the model), D$_\mathrm{m,9}$ is the diffusion coefficient for methyl-group rotation in rotamers 9 and $\alpha_\mathrm{CSA}$ is a scaling factor applied to the rotamer-specific CSA amplitudes. The values of p$_i$ obtained from the MD simulation are shown with the red lines while mean and width of the distributions are shown with the orange solid and dash vertical lines respectively.}
\end{figure*}

\begin{figure}
\begin{center}
\includegraphics[width=0.9\textwidth]{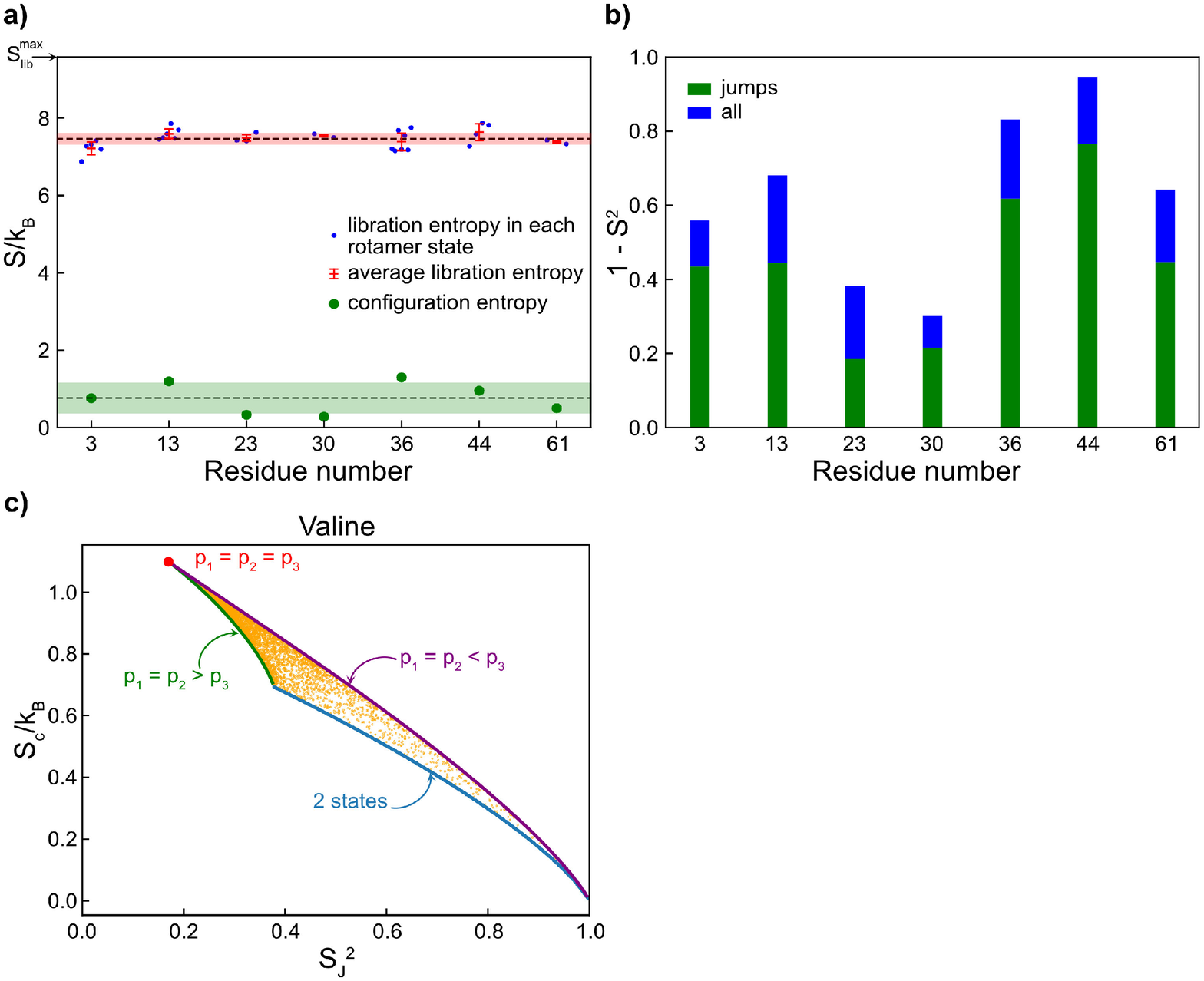}
\end{center}
\caption{\label{fig:EntropyMD}
a) Entropy obtained from the molecular dynamics simulation. The libration entropy is defined within each rotamers as $S_{lib} = \int_{\chi_1 - 60}^{\chi_1 +60} \int_{\chi_2 - 60}^{\chi_2 +60} p(x, y) \ln p(x, y) dx dy$ with $\chi_1$ and $\chi_2$ the dihedral angles defining the rotamer of interest. This entropic term only included contribution from CC bond diffusion within a rotamer state. The entropy $S_{lib}$ is shown for rotamers populated more than 1\%, with the average and standard deviation with the red crosses. Note that the standard for the value of $S_{lib}$ depends on the choice of units. Thus only variations in $S_{lib}$ are significant, not its absolute value. The configuration entropy is shown in green circles. The red and green areas represent the ranges for the libration and configuration entropies, respectively, over all residues. These areas are centered on the average values shown with horizontal dashed lines, and have widths corresponding to the standard deviation calculated over all residues. The maximum value for the libration entropy within a rotamer state is indicated on the left y-axis and is calculated as $S_\mathrm{lib}^\mathrm{max} = - \int_{-60}^{+60} \int_{-60}^{+60} p(\chi_1)  p(\chi2) \ln (p(\chi_1)  p(\chi2)) d\chi_1 d\chi_2 = 2 \ln 120$ for a uniform distribution of the $\chi_1$ and $\chi_2$ dihedral angles and integration step of 1$^\circ$.  Integration over ranges of 120$^\circ$ allows us to span the entire $\{\chi_1$, $\chi_2\}$ surface for each of the 9 rotameric states. b) Value of $1 - \mathcal{S}^2$ as calculated from the MD trajectory. The contribution of the rotamer jump is shown with the green bar. The value of $1 - \mathcal{S}^2$ indicates the amplitude of motions. c) Configuration entropy ($S_c$) and squared order parameter for rotamer jump ($S_J^2$) for 10,000 random rotamer distribution of a valine side-chain with ideal geometry. The conditions on the populations leading to the borders of the area spanned by the sampling are indicated on the figure. The numbering of rotameric states is irrelevant as ideal $\mathcal{C}_3$ geometry is assumed for rotamer states.}
\end{figure}

Our approach is illustrated in the case of carbon-13 relaxation in isoleucine-$\delta$1 specifically labeled $^{13}$C$^1$H$^2$H$_2$ methyl groups, where carbon-13 relaxation is driven by its dipole-dipole (DD) interaction with the proton and deuterons, and its chemical shift anisotropy (CSA).  We analyze a broad dataset consisting of carbon-13 longitudinal (R$_1$) and transverse (R$_2$) relaxation rates, carbon-proton DD cross-relaxation rates ($\sigma^\mathrm{NOE}$) measured at four high magnetic fields, as well as 22 relaxometry carbon-13 R$_1$ rates measured over two orders of magnitude of magnetic field \cite{Cousin_JACS_2018}.  We also include CSA/DD cross-correlated cross-relaxation (CCCR) rates (longitudinal $\eta_\mathrm{z}$ and transverse $\eta_\mathrm{xy}$) measured at two high magnetic fields.  We use for all models a Markov-Chain Monte-Carlo (MCMC) procedure \cite{Emcee2013} to analyze 38 relaxation rates per methyl group and extract the model parameters. \\
We previously used an MF-type of correlation function to describe the isoleucine C$_{\gamma1}$C$_{\delta1}$ bond motions, \textit{i.e.} the extended model-free \cite{Clore_JACS_1990} correlation function. When all rates are analyzed together, experimental transverse relaxation rates R$_2$ cannot be reproduced for any of the six isoleucines residues with no sizeable contribution of chemical exchange, especially at the highest magnetic fields (Fig.\,\ref{fig:Fit61}.a and Fig.\,\ref{fig:MFfits}). These large discrepancies demonstrate that MF-type of correlation functions cannot reproduce this set of relaxation rates. \\
\begin{figure}
\begin{center}
\includegraphics[width=0.5\textwidth]{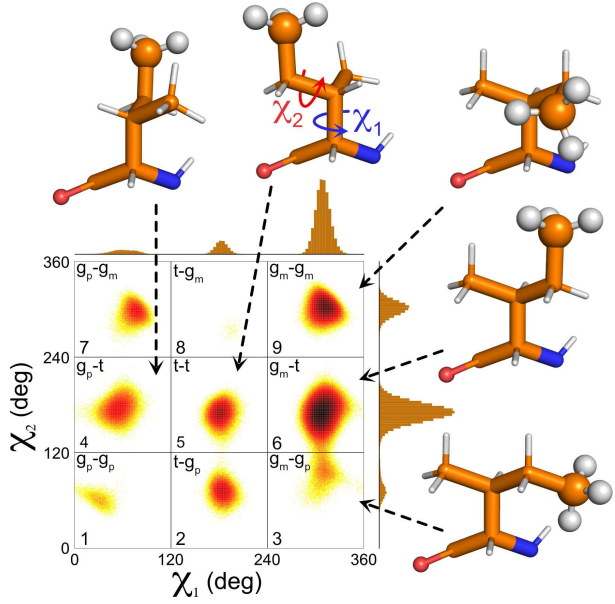}
\caption{\label{fig:Ramachandran}
Probability density distribution for $\chi_1$ and $\chi_2$ dihedral angles for Ile-36 obtained from a 1\,$\mu$s MD trajectory.  Nine rotamers can be defined, each numbered from 1 to 9 as shown. We also indicate the common states nomenclature with \textit{gauche+} (g$_\mathrm{p}$), \textit{trans} (t) and \textit{gauche-} (g$_\mathrm{m}$). Some conformers are shown as well with the C$_{\delta1}$H$_3$ group presented with spheres.}
\end{center}
\end{figure}
Recent advances in MD simulations, with improved protein force fields, provide a wealth of information on the nature of motions \cite{Cousin_JACS_2018, Kummerer_JCTC_2021, Smith_JMRO_2022}. We propose to use an MD trajectory to define appropriate explicit models of correlation functions to analyze experimental data, providing a mechanistic description of motions. Note that we are not: (i) validating one method with the other, as can be done by calculating NMR parameters from MD simulations \cite{Maragakis_JPCB_2008,Kummerer_JCTC_2021}; (ii) constraining  an MD trajectory using NMR results \cite{Lindorff_nature_2005,Camilloni_JACS_2012}; (iii) reweighing an MD trajectory with NMR constraints \cite{Kummerer_JCTC_2021}. Rather, we use MD to identify the rotamers relevant for an explicit model that we use to analyze NMR relaxation. Our approach echoes a recent study of lipid bilayer dynamics \cite{Smith_NatComm_2022} where MD results were refined using NMR data, but the detectors approach used therein \cite{Smith_NatComm_2022, Smith_JCP_2018} yielded no mechanistic description of motions.  \\
We performed a 1\,$\mu$s MD simulation of Ubiquitin with Gromacs \cite{Berendsen_CPC_1995,Lindahl_JMolModel_2001,Spoel_JCC_2005,Hess_JCTC_2008,Pronk_BioInf_2013}, using the Amber ff99SB*-ILDN force field \cite{Best_JPCB_2009,Lindorff_Proteins_2010} modified with accurate energy barriers for methyl rotation \cite{Hoffmann_JPCB_2018} and the TIP4P-2005 water model \cite{Abascal_JCP_2005}. The distributions of the dihedral angles $\chi_1$ and $\chi_2$ highlight the different possible motions of each isoleucine residue (Fig.\,\ref{fig:Ramachandran} and Fig.\,\ref{fig:MDdistributions}). Accessible conformations of isoleucine side-chains correspond to the 9 possible ($\chi_1$, $\chi_2$) rotamers and their dynamics is well described by instantaneous jumps between rotamer states.  Here, we neglect the local librations of C$_{\gamma1}$C$_{\delta1}$ bonds as we assume a negligible contribution to relaxation \cite{BolikCoulon_ArXiv_2021}, and use a model of infinitely fast jumps between discrete positions for C$_{\gamma1}$C$_{\delta1}$ bond dynamics \cite{Wittebort_JCP_1978}. \\
The chemical environment for the Ile-$\delta$1 methyl-groups is rotamer-dependent, which should lead to rotamer-specific chemical shift tensors \cite{Siemons_ChemCom_2019}. Precise isoleucine side-chain conformations in water were obtained using Density Functional Theory (DFT) calculations in Gaussian \cite{Gaussian09} (see supplementary information). The CSA tensors calculated with the Gauge-Independent Atomic Orbital method \cite{Zeroka_JCP_1966,Ditchfield_JCP_1972} are rotamer dependent (Table\,\ref{table:CSAtensors}). Within the Born-Oppenheimer approximation, any jump of the side chain from one rotamer to another would thus result in an instantaneous change of the CSA. We included this time-dependent interaction amplitude in our model to accurately account for the fluctuations of CSA tensors. \\
In the Bloch-Wangsness-Redfield relaxation theory, the correlation function between interactions \textit{i} and \textit{j} can include the amplitudes of the interactions, as denoted by the superscript (I):
\begin{figure*}
\includegraphics[width=0.95\textwidth]{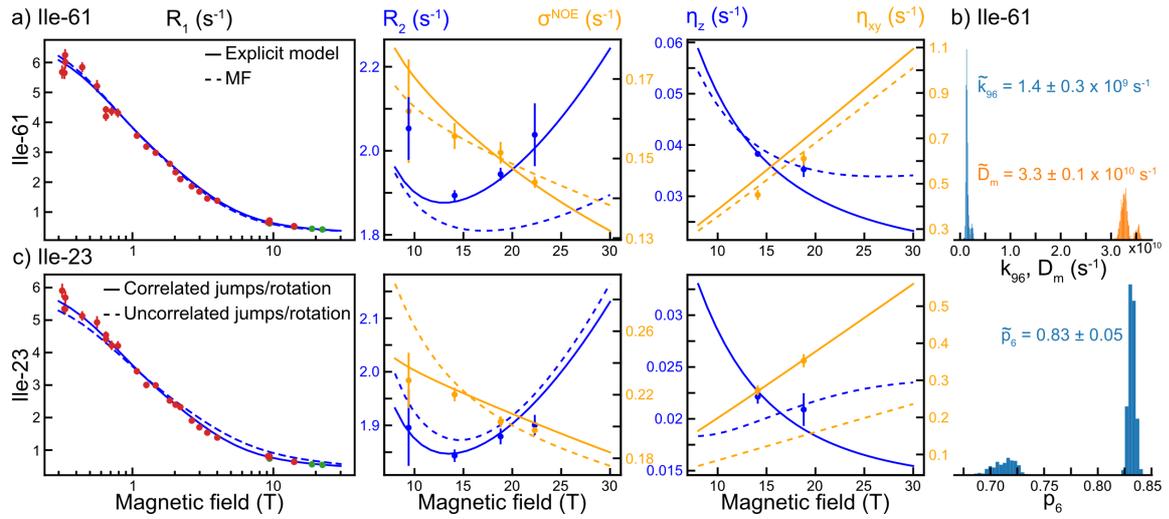}
\caption{\label{fig:Fit61}
Explicit models of motions for residues Ile-61 and Ile-23 in human Ubiquitin were used to analyze longitudinal $R_1$ and transverse $R_2$ relaxation rates; dipolar cross-relaxation rates $\sigma^{NOE}$; longitudinal $\eta_{z}$, and transverse $\eta_{xy}$ CCCR rates. a) Measured and calculated relaxation rates following a MCMC procedure using the explicit model of motions (solid line) or MF-type of correlation function (dashed line) for Ile-61. b) Distributions of parameters of the explicit model of motions for Ile-61 from the MCMC analysis. MD simulations show that only rotamers 6 and 9 are populated, suggesting a model including jumps between rotamers 6 and 9. The free parameters of the model are the population of rotamer 6, $p_6$, the exchange rate from rotamer 6 to rotamer 9, $k_{96}$, and the diffusion constant for rotation on a cone of the methyl group, $D_m$.  The median and standard deviation are indicated on each panel. c) Measured and calculated relaxation rates following a MCMC procedure using the explicit model of motions with (solid line) or without (dashed line) correlation of jumps and methyl rotation for Ile-23. The high-field R$_1$ and the corrected relaxometry rates are shown in green and red respectively.}
\end{figure*}
\begin{equation}
C_{i,j}^{(I)}(t) = \langle \zeta_i (0) \zeta_j(t) \mathcal{D}_{q0}^{(2)*} (\Omega_{L,i}, 0) \mathcal{D}_{q0}^{(2)*} (\Omega_{L,j}, t) \rangle,
\end{equation}
where $\langle ... \rangle$ stands for an ensemble average, $\mathcal{D}_{q0}^{(2)*} (\Omega_{L,i}, t)$ is a rank-2 Wigner matrix with the Euler angle set $\Omega_{L,i} = \{\alpha_{L,i}, \beta_{L,i}, \gamma_{L,i}\}$ for transformation from the laboratory to the interaction-$i$ frame at time \textit{t}, and $\zeta_i(t)$ is the strength of the interaction \textit{i} at time \textit{t}. Motions are modeled using Fokker-Plank diffusion equations and the associated operators are diagonalized to write time-dependent bond-orientation conditional probabilities used to express the correlation function \cite{Luginbuhl_PNMR_2002,BolikCoulon_ArXiv_2021}. Intermediate frames are introduced to facilitate the description of individual motions, all assumed to be statistically independent unless otherwise stated (Fig.\,\ref{fig:FrameTransformation}): the global tumbling (with the diffusion frame), the rotamer jump (with the jump and rotamer frames) and the methyl rotation (with the system frame). A diffusing motion for the methyl rotation is a relevant approximation, as discussed in section I of the supplementary information. Note that both internal and global motions are anisotropic \textit{a priori}. \\
The complexity of the correlation functions rapidly increases with the number of rotamers $N$ as the sums contain $N^3$ terms, and $N(N-1)/2$ jump rates need to be determined to fully characterize the exchange matrix. Information from MD simulations is essential to define suitable explicit models with a minimal number of adjustable parameters. First, we only included rotamers with average fractional populations higher than 1\,\% in the MD trajectory (Table\,\ref{table:PopMD}). Second, jump rates for transitions which were not observed in the MD trajectory were fixed to 0 in the exchange matrix. The adjustable parameters of the model are populations of rotamers selected by MD, exchange rates and diffusion constants for methyl rotation. \\
Explicit models of motions are compatible with experimental relaxation rates (Fig.\,\ref{fig:Fit61}.a and Fig.\,\ref{fig:FitsAll}) with better agreement than the MF analysis (Table\,\ref{table:Chi2All}), in particular for R$_2$ and CCCR rates. Some deviations can be observed for the $\sigma^\mathrm{NOE}$, which might be due to the neglected contribution of fast C$_{\gamma1}$C$_{\delta1}$ bond librations \cite{BolikCoulon_ArXiv_2021}. In addition, the explicit model gives a mechanistic picture of the motion: we estimate equilibrium populations (Fig.\,\ref{fig:Fit61}.b, Fig.\,\ref{fig:PopDistributions} and Table\,\ref{table:PopRelax}) which are only accessible to a few advanced experimental methods \cite{Hansen_JACS_2010,Hansen_JACS_2011,Siemons_ChemCom_2019} and, uniquely, the kinetics of exchange. \\
Most distributions of parameters obtained from the MCMC analysis are well defined and allow a precise estimate of the parameters of the model (Eq.\,\ref{eq:AllCorrFunc}-\ref{eq:CorrelatedJumpMatrix}, Fig.\,\ref{fig:Fit61}.b, Fig.\,\ref{fig:PopDistributions}, \ref{fig:DynDistributions}). However, as the number of states increases, the distributions become broader and the resulting jump matrices can be ill-defined (see supplementary information). Further improvement could consist in analyzing relaxation rates simultaneously with other data defining conformational ensembles, such as chemical shifts or scalar-coupling constants \cite{Chou_JACS_2003,Siemons_ChemCom_2019}. \\
Explicit models of motions constitute an efficient framework to combine the imperfect information from MD simulations and experiments. Despite recent force field improvements \cite{Hoffmann_JPCB_2018}, MD data alone cannot reproduce relaxation rates, whether these are calculated directly from the MD correlation functions (Fig.\,\ref{fig:MDcorrFuncFits}, \ref{fig:MDrates_MF}) or from the analysis of MD with the explicit models of motions \cite{Beauchamp_JCTC_2011} (Fig.\,\ref{fig:MDrates_Explicit} and Table\,\ref{table:Chi2All}). Beyond force-field deficiencies, this is likely due to the limited sampling on the microsecond timescale. If the identification of the relevant rotameric states is probably reliable, the extraction of accurate transition rates would require many more transitions during the MD trajectory for good statistics. \\
Nevertheless, the MD input is critical to complement the information from experimental relaxation rates. We analyzed relaxation rates for isoleucines 30 and 61 for all combinations of two rotameric states.  Several sets of two rotamers lead to comparable or slightly better $\chi^2$ values (Fig.\,\ref{fig:61DifRot}) than the combination of rotamers obtained from the MD simulation. Such an exhaustive analysis is challenging for a two-state model even assuming prior knowledge on the number of states. For larger numbers of states, such an approach would be unrealistic. 
Thus, relaxation rates alone are insufficient to determine the populations and kinetics of exchange among rotamer states. Our approach combines the most robust  information from an MD simulation: the network of accessible rotamer states, with the information from NMR relaxation, which is sensitive to the populations of rotamer states and the kinetics of exchange. This combined analysis of MD simulation and NMR is necessary to obtain a quantitative mechanistic description of the dynamics. \\
Explicit models of motion can accommodate motions of increasing complexity including, for instance, different diffusion rates for methyl rotation in rotamer states. In the case of Ile-23, the explicit model of motions with uniform methyl rotation (Eq.\,\ref{eq:AllCorrFunc}) is unable to reproduce the experimental data well (Fig.\,\ref{fig:FitsAll}). We noticed that for the major rotamer conformation of this residue, the $\delta1$-methyl group is in close proximity of the H$_\alpha$ (rotamer 9 in Fig.\,\ref{fig:Ramachandran}). Such steric hindrance is not present in the other rotamer states. We built a model taking into account methyl rotation specific to rotamer states, with identical diffusion constants for methyl rotation in rotamers 3 and 6, and a specific diffusion constant for rotamer 9, which led to a clear improvement in the agreement of the model with the measured relaxation rates (Fig.\,\ref{fig:Fit61}.c and Fig.\,\ref{fig:Fit23}). The resulting diffusion constants for  methyl rotation $D_{m,3} = D_{m,6} = 5.5 \pm 1.9 \times 10^{10}$\,s$^{-1}$ and $D_{m,9} = 1.1 \pm 0.4 \times 10^{10}$\,s$^{-1}$ support the presence of correlated motions in Ile-23 side chain, with  methyl rotation 5 times slower in the major rotamer, as expected from the steric hindrance of the H$_\alpha$. \\
Generalized order parameters quantify the width of the conformational space at equilibrium, and can thus be linked to conformational entropy \cite{Akke_JACS_1993,Frederick_nature_2007}, which is defined by considering each accessible conformation as a microstate. The link between the generalized order parameter obtained in the MF approach and entropy is not direct and requires either to reintroduce models of motion or rely on residue-specific estimates from MD simulations \cite{Akke_JACS_1993,Yang_JMB_1996,Li_JACS_2009}. All these models have to use a single parameter to describe the amplitude of motions. One may ask whether a single parameter suffices to describe both order parameters and conformational entropy. In stark contrast to existing approaches, our explicit models of motions provide the distribution of rotamer states that can be directly used to estimate the configuration entropy, $S_c$ associated to this distribution of rotamer states:
\begin{equation}
S_c/k_B = -\sum_{\alpha=1}^N p_\alpha \ln p_\alpha,
\label{eq:confEntropy}
\end{equation}
where $k_B$ is the Boltzmann constant and $p_\alpha$ is the equilibrium fractional population of rotamer $\alpha$. The full conformational entropy includes both the configuration entropy $S_c$ and the differential entropy \cite{Singh_AJMMS_2003} associated with substates within each rotamer and quantified from the amplitude of librations. The simulation analysis shows that order parameters predominantly reflect the equilibrium distribution of rotamers, suggesting a close link between order parameters and configuration entropy (Fig.\,\ref{fig:EntropyMD}.b). Importantly, we find that the differential entropy in a given rotameric state is rather constant, between rotamers and between residues (Fig.\,\ref{fig:EntropyMD}.a). Thus, variations in conformational entropy mostly arise from configuration entropy. The rotamer distributions enable to evaluate the relationship between order parameters and configuration entropy. We generated 10,000 random  distributions of rotamer populations, using optimized geometries for side-chains obtained from DFT calculations (see above and supporting information), and computed both configuration entropy (Eq.\,\ref{eq:confEntropy}) and order parameters according to \cite{Chou_JACS_2003,BolikCoulon_ArXiv_2021}:
\begin{equation}
\mathcal{S}_J^2 = \sum_{\alpha,\beta=1}^N p_\alpha p_\beta \mathcal{P}_2(\cos \theta_{\alpha,\beta}),
\label{eq:S2J}
\end{equation}
where $\theta_{\alpha,\beta}$ is the angle between rotamers $\alpha$ and $\beta$. The random distributions were generated by first randomly choosing the number of populated rotamers, and then the fractional population for each of them. \\
\begin{figure}
\begin{center}
\includegraphics[width=0.5\textwidth]{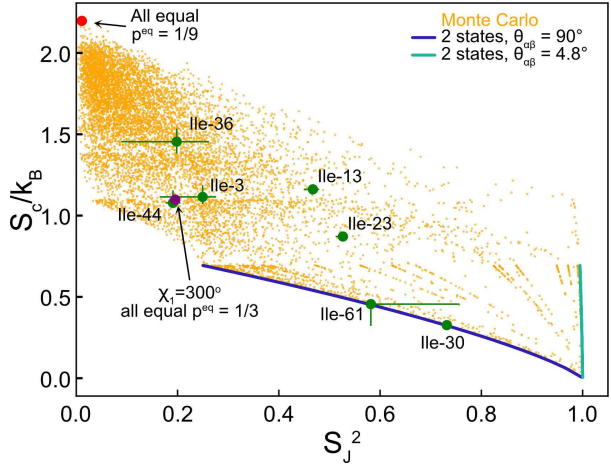}
\caption{\label{fig:Entropy}
Configuration entropy ($S_c$, Eq.\,\ref{eq:confEntropy}) and squared order parameter for rotamer jump ($S_J^2$, Eq.\,\ref{eq:S2J}) calculated for 10,000 random rotamer distribution of an isoleucine side-chain with geometry derived from DFT calculations. The generalized order parameters and configuration entropies obtained from the combined analysis of NMR relaxation and MD simulations for the 7 isoleucines of Ubiquitin are shown in green, with errors corresponding to the 16$^\mathrm{th}$ and 84$^\mathrm{th}$ percentile of the distributions following calculations of the order parameters and configuration entropy over the whole MCMC trajectories. The blue and green parametric curves describe a two-rotamer jump model with the lowest (blue) and highest (cyan) value for $\mathcal{P}_2 (\cos \theta_{\alpha,\beta})$. The point where all rotamers have equal population (\textit{i.e.} highest entropy) is shown in red and the point where three rotamers have equal population and $\chi_1=300^\circ$ (rotamers 3, 6 and 9) is shown in purple.}
\end{center}
\end{figure}
Interestingly, for a given value of order parameter, we find a range of possible configuration entropies as wide as \textit{ca.} $k_B$  (Fig.\,\ref{fig:Entropy}): configuration entropy cannot be derived unambiguously from an order parameter. Similar results are obtained for leucine and valine side-chains (Fig.\,\ref{fig:EntropyMD}.c).  Our combined NMR and MD analysis gives distributions of rotamer populations that can be used to derive both configuration entropies and order parameters. Importantly, our experimental results confirm the calculations from random distributions (Fig.\,\ref{fig:Entropy} and Table\,\ref{table:Entropie}): large variations of configuration entropy can be observed for small changes of order parameters (see Ile-13 and Ile-61) and similar configuration entropies can be obtained for residues with drastically different order parameters (compare Ile-13 and Ile-44).  Considering additional limitations arising from limited conformational sampling on the timescale of overall rotational diffusion \cite{Hoffmann_JPCB_2022}, our investigation confirms that the estimation of conformational entropy from NMR data alone is difficult. The combined analysis of NMR relaxation and MD simulations with a rotamer jump model uncovers the physical origin of conformational entropy and solves limitations that arise when a single-parameter model is used to link order parameters and conformational entropy. Further integration may open a path towards the identification of correlated motions in systems with a larger number of experimental side-chain probes \cite{Tugarinov_NatProtoc_2006,Gans_ANIE_2010}. \\

Here, we have shown that molecular dynamics simulations could be used to build explicit models of motions in order to obtain a mechanistic description of motions for protein side-chains from the analysis of NMR relaxation. We defined models of motions from MD simulations that include jumps between rotamer states and methyl rotation on a cone. We determined from NMR relaxation the populations of rotamers and their exchange rates. Such explicit models allow fine-tuning of motional properties, such as slower methyl rotation in one rotamer state, which may be understood by steric interactions. The description of motions as exchange between substates gives access to the molecular origin of conformational entropy, the variations of which are dominated by the configuration entropy associated to the distribution of rotamer states. Our approach, based on a combination of DFT calculations, MD simulations, and extensive NMR measurements can be adapted for most probes of site-specific dynamics in macromolecules. We anticipate that further integration of NMR and MD simulations in combination with explicit models of motion will improve the mechanistic description of protein motions and lead to a better understanding of the physics and chemistry that sustain protein function. \\

\textbf{Acknowledgments} The project was supported by the grant ALLODYN from the PSL Chemistry call of PSL University, as well as, FET-Open project of the European Union, Grant agreement 899683 (HIRES-MULTIDYN). This work was also supported by the "Initiative d'Excellence" program from the French State (Grant "DYNAMO", ANR-11-LABX-0011-01, GS and FS), and simulations benefited from the access to the HPC resources of TGCC under the allocation A0070811005 made by GENCI (Grand Equipement National de Calcul Intensif) to GS. \\


\bibliographystyle{ieeetr}

\bibliography{Biblio}

\clearpage
\newpage

\section*{Supplementary information}
\beginsupplement

\section{Correlation functions}
\paragraph{Model-Free type of correlation functions}
We use the following expression of correlation function when using a MF-type of analysis [1]:
\begin{equation}
C_{MF}(t) = C_g(t) C_{CC}(t) C_{m}(t, i, j)
\end{equation}
where we assume that the rotational diffusion, the motions of the C$_{\gamma1}$-C$_{\delta1}$ bond and the methyl group rotation are independent, each of them described by the individual correlation functions $C_g$, $C_{CC}$ and $C_{m}$ respectively:
\begin{subequations}
\label{eq:AllCorrFuncEMF}
\begin{equation}
C_g (t) = \frac{1}{5} e^{-t/\tau_c},
\end{equation}
\begin{equation}
C_{CC}(t) = \mathcal{S}_f^2 \mathcal{S}_s^2 + (1 - \mathcal{S}_f^2)e^{-t/\tau_f} + \mathcal{S}_f^2 (1 - \mathcal{S}_s^2) e^{-t/\tau_s},
\end{equation}
\begin{equation}
C_{m}(t, i, j) = \mathcal{S}_{m}^2(i,j) + (\mathcal{P}_2(\cos \theta_{i,j}) - \mathcal{S}_{m}^2(i,j)) e^{-t/\tau_{m}},
\end{equation}
\end{subequations}
where $\tau_c$ is the correlation time for global tumbling, $\tau_f$ and $\tau_s$ are correlation times for fast and slow motions of the C$_{\gamma1}$-C$_{\delta1}$ bond, associated to the order parameters $\mathcal{S}_f$ and $\mathcal{S}_s$ respectively, $\tau_{m}$ is the correlation time for methyl rotation, $\mathcal{P}_2(x) = (3x^2 -1 )/2$ is the second order Legendre polynomial, $\theta_{i,j}$ is the angle between the main axis of the interactions $i$ and $j$ frames, and the squared order parameter for methyl rotation is defined as $\mathcal{S}_m^2 (i,j) = \mathcal{P}_2(\cos \theta_{CC,i}) \mathcal{P}_2(\cos \theta_{CC,j})$ with $\theta_{CC,k}$ the angle between the C$_{\gamma1}$-C$_{\delta1}$ bond and the principal axis of the interaction \textit{k}. In this MF-type of analysis, we assume that the overall tumbling is isotropic with correlation time $\tau_c$ equal to 5.028 ns [1], and that the $^{13}$C-CSA frame is aligned along the C$_{\gamma1}$-C$_{\delta1}$ bond. \\

\paragraph{Explicit models of motions}
In order to write the correlation functions for explicit models of motion, we solve Fokker-Planck equations, which we report below for the motions considered in this study.
\begin{itemize}
\item For the global tumbling:
\begin{equation}
\frac{\partial}{\partial t} P(\Omega_{Lab,D}, t) = - \sum_{j=x,y,z} D_{jj} L_j^2 P(\Omega_{Lab,D},t),
\label{eq:FPglobaltumbling}
\end{equation}
where $\Omega_{Lab,D}$ is the Euler angle for transformation from the laboratory to diffusion frame, $P(\Omega_{Lab,D}, t)$ is the probability to find angle $\Omega_{Lab,D}$ at time $t$, $D_{jj}$ is the $j$th component of the diagonalized diffusion tensor and $L_j$ is the associated angular momentum operators. 
\item For the rotamer jumps:
\begin{equation}
\frac{\partial}{\partial t}p_\alpha (t)  = \sum_{\beta=1}^N \mathcal{R}_{\alpha \beta} p_\beta (t),
\label{eq:MasterEqJump}
\end{equation}
where $p_\alpha (t)$ is the population of state $\alpha$ at time $t$ and $\mathcal{R}_{\alpha \beta}$ is the exchange rate from state $\beta$ to $\alpha$. 
\item For the rotation on a cone:
\begin{equation}
\frac{\partial}{\partial t} p(\gamma, t) = - D_m L_{rot}^2 p(\gamma, t) = D_m \frac{\partial^2}{\partial \gamma^2} p(\gamma, t),
\label{eq:MasterEqRot}
\end{equation}
where $D_m$ is the rotational diffusion coefficient, $\gamma$ is the angle characterizing the rotation and $p(\gamma, t)$ is the probability to find angle $\gamma$ at time $t$.
\item For correlated rotamer jumps and rotation on a cone:
\begin{equation}
\frac{\partial}{\partial t}p_\alpha (\gamma, t)  = \sum_{\beta=1}^N \mathcal{R}_{\alpha \beta} p_\beta (\gamma, t) + D_{m,\alpha} \frac{\partial^2}{\partial \gamma^2} p_\alpha(\gamma, t),
\label{eq:MasterEqJump}
\end{equation}
with the same notations as above, and with $p_\alpha (\gamma, t)$ the probability to find the system in rotamer $\alpha$ with rotation angle $\gamma$ at time $t$ and $D_{m,\alpha}$ the rotational diffusion coefficient in rotamer $\alpha$.
\end{itemize}

\begin{figure}[!ht]
	\begin{center}
		\includegraphics[width=0.9\textwidth]{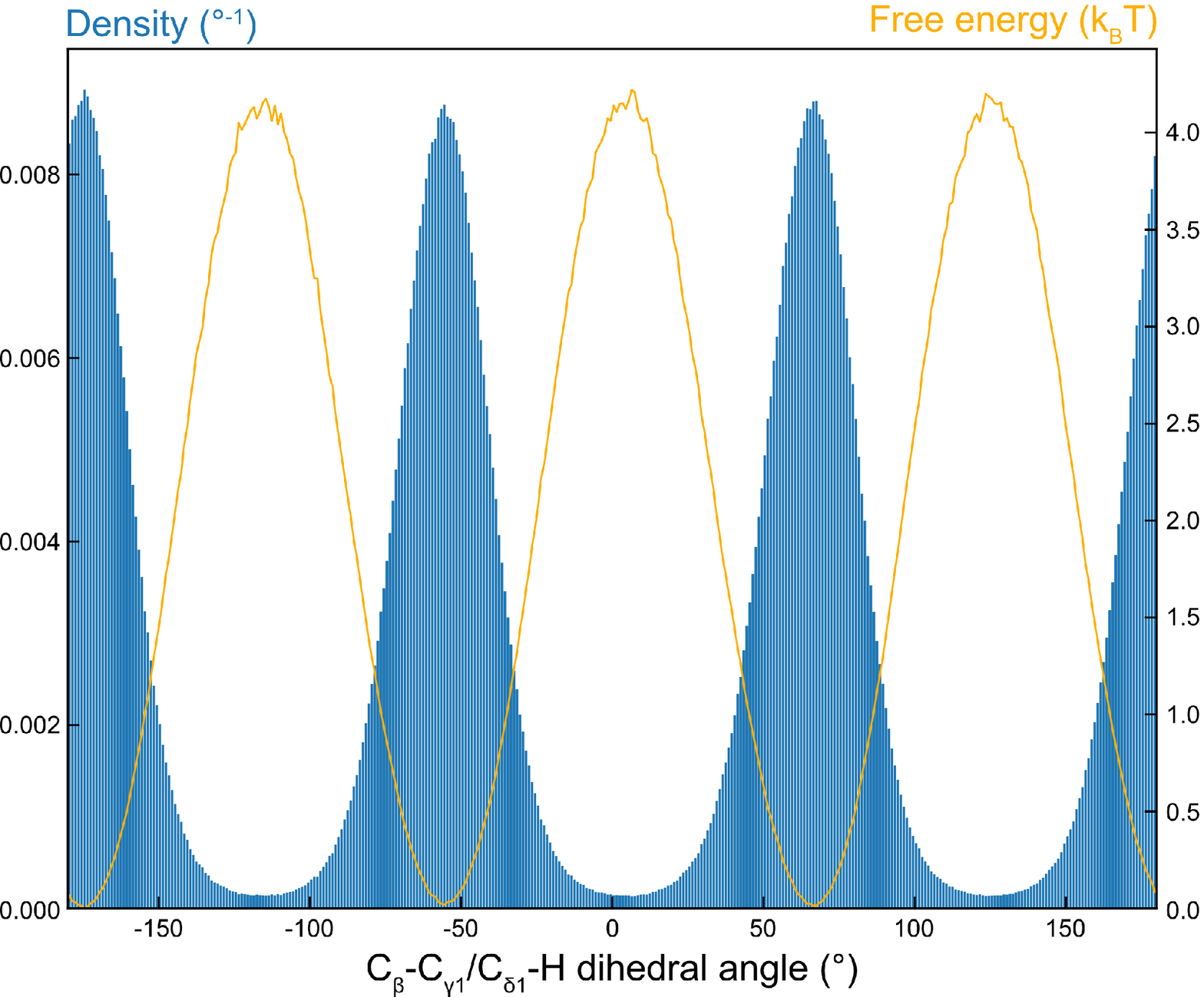}
	\end{center}
	\caption{Distribution of dihedral angles associated to the methyl rotation obtained from the MD simulation, and associated free energy in units of $k_BT$, with $k_B$ the Boltzmann constant and $T$ the temperature. Results are shown for Ile-36.}
	\label{fig:MethylRotMD}
\end{figure}

The correlation functions for explicit models of motions accounting for rotamer jumps, methyl rotation and overall rotational diffusion with an axially symmetric diffusion tensor are:
\begin{subequations}
\label{eq:AllCorrFunc}
\begin{equation}
\begin{aligned}
C_{CX}^{(I)}(t) =  \frac{d_{CX}^2}{5} \sum_{a,b=-2}^2 e^{-(6 D_\perp + a^2 (D_\parallel - D_\perp))t}  \sum_{\alpha,\beta,n=1}^N \sqrt{p_\alpha p_\beta} \tilde{X}_\alpha^{n} \tilde{X}_\beta^{n} e^{(\lambda_n - b^2 D_m) t} \times & \\
	\mathcal{D}_{a,b}^{(2)*} (\Omega_{D,R_\alpha}) \mathcal{D}_{a,b}^{(2)} (\Omega_{D,R_\beta})  d_{b0}(\beta_m)^2, &
\end{aligned}
\end{equation}
\begin{equation}
\begin{aligned}
C_{\sigma_i,\sigma_j}^{(I)} (t) =  \frac{2\omega_C^2}{15}  \sum_{a=-2}^2 e^{-(6 D_\perp + a^2 (D_\parallel - D_\perp))t}  \sum_{\alpha,\beta,n=1}^N \Delta \sigma_i^{\alpha} \Delta \sigma_j^{\beta} \sqrt{p_\alpha p_\beta} \tilde{X}_\alpha^{n} \tilde{X}_\beta^{n} e^{\lambda_n t}  \times & \\
	\mathcal{D}_{a,0}^{(2)*} (\Omega_{D,\sigma_i^{\alpha}}) \mathcal{D}_{a,0}^{(2)} (\Omega_{D,\sigma_j^{\beta}}) , &
\end{aligned}
\end{equation}
\begin{equation}
\begin{aligned}
C_{\sigma_i,CH}^{(I)} (t) = \sqrt{\frac{2}{3}}  \frac{d_{CH}  \omega_C }{5} \mathcal{P}_2(\cos \beta_m)\sum_{a=-2}^2 e^{-(6 D_\perp + a^2 (D_\parallel - D_\perp))t}  \sum_{\alpha,\beta,n=1}^N \Delta \sigma_i^{\alpha} \sqrt{p_\alpha p_\beta} \tilde{X}_\alpha^{n} \tilde{X}_\beta^{n} e^{\lambda_n t} \times & \\
	 \mathcal{D}_{a,0}^{(2)*} (\Omega_{D,\sigma_i^{\alpha}}) \mathcal{D}_{a,0}^{(2)} (\Omega_{D,R_\beta}), &
\end{aligned}
\end{equation}
\end{subequations}
where $C_{CX}^{(I)}$ is the auto-correlation function for DD interactions ($X=H,D$), $C_{\sigma_i,\sigma_j}^{(I)}$ is the correlation function for CSA interactions with $i$ and $j$ referring either to the longitudinal or orthogonal component of the CSA tensors and $C_{\sigma_i,CH}^{(I)}$ is the CSA/DD cross-correlation function. The definition of each parameter is gathered in Table\,\ref{table:ParamDef}. \\
When we considered correlated rotamer jumps and methyl rotation, only the DD auto-correlation function is modified to:
\begin{equation}
\label{eq:CorrFuncCorrelatedJumpsRot}
\begin{aligned}
C_{CX}^{(I,\mathrm{corr})}(t) =  \frac{d_{CX}^2}{5} \sum_{a,b=-2}^2 e^{-(6 D_\perp + a^2 (D_\parallel - D_\perp))t}  \sum_{\alpha,\beta,n=1}^N \sqrt{p_\alpha p_\beta} \tilde{X}_\alpha^{n} \tilde{X}_\beta^{b,n} e^{\lambda_{b,n} t} \times & \\
\mathcal{D}_{a,b}^{(2)*} (\Omega_{D,R_\alpha}) \mathcal{D}_{a,b}^{(2)} (\Omega_{D,R_\beta})  d_{b0}(\beta_m)^2, &
\end{aligned}
\end{equation}
where $\lambda_{b,n}$ and $\tilde{X}^{(b,n)}$ are eigenvalues and eigenvectors for the symmetrized pseudo-jump matrix defined by:
\begin{equation}
\label{eq:CorrelatedJumpMatrix}
\tilde{\mathcal{R}}_{b} = \tilde{\mathcal{R}} - b^2 \mathcal{D},
\end{equation}
where $\mathcal{D}$ is a diagonal matrix containing the methyl rotation diffusion coefficients as diagonal elements. \\
The overall diffusion and orientations of the side-chains in the diffusion tensor frame were determined with RotDiff [2] using backbone $^{15}$N-$^1$H relaxation data recorded in H$_2$O [3]. After accounting for the difference in viscosity of H$_2$O and D$_2$O, we obtained $D_\parallel = 3.99 \times 10^7$\,s$^{-1}$ and $D_\perp = 3.39 \times 10^7$\,s$^{-1}$ for the specifically labelled [$^{13}$C$^1$H$^2$H$_2$] Ile-$\delta1$ Ubiquitin. The orientation of the rotamer frames in the jump frame, and of the CSA tensor frames in the rotamer frames were obtained from the DFT calculations.  \\
\paragraph{Discussion on the model for methyl rotation} In our models of correlation functions, we assume the methyl rotation to be diffusive (all orientations of the C$_{\delta1}$-H bonds on the cone surface are equiprobable). A model of 3-site jump model could also be used to represent this type of motion (Fig.\,\ref{fig:MethylRotMD}). It must be noted that even if the jump model seems physically relevant, these two options are limit models since the C$_{\delta1}$-H bond also diffuses around each of its three preferred orientations. Here, methyl-rotation correlation times are all smaller than 20\,ps (with the exception of rotation in rotamer 9 for Ile-23) (Table\,\ref{table:MetRotCorrTimes}) in which case we have shown that both models lead to identical relaxation rates [4]. This is also confirmed by the distributions of the C$_{\beta}$-C$_{\gamma1}$/C$_{\delta1}$-H dihedral angle (Fig.\,\ref{fig:MethylRotMD}) obtained from the MD simulation: the energy barier from one orientation to the other is of the order of $4k_BT$ ($k_B$ the Boltzmann constant and $T$ the temperature) such that the residence time estimated from transition state theory is \textit{c.a.} 8\,ps, orders of magnitude below the typical transitions for $\chi_1$ and $\chi_2$ transitions. Thus, the free-energy landscape associated to the rotation can be considered flat with a roughness related to the preferred orientations of the C$_{\delta1}$-H bonds. In such cases, and for the range of temperatures relevant in the study of biomolecules, diffusion can be treated with an effective diffusion coefficient [5]. Such approximation might not hold for other types of side-chain, but explicit models of motions are flexible enough to treat methyl rotation as a 3-site jump in a straightforward manner [4].

\section{Expression of relaxation rates}
We report here the expression of the relaxation rates for a $^{13}$C$^1$H$^2$H$_2$ methyl groups, as a function of the spectral density functions and for an axially symmetric carbon-13 CSA tensor aligned with the methyl axis:
\begin{subequations}
\label{eq:AllRatesMF}
\begin{eqnarray}
\mathrm{R}_1 =&& \frac{1}{3} \Delta \sigma_c^2 \omega_C^2 \mathcal{J}_C (\omega_C) + \nonumber \\
		&&  \frac{1}{4} d_{CH}^2 \left(\mathcal{J}_{CH} (\omega_C - \omega_H) +  3 \mathcal{J}_{CH} (\omega_C) + 6 \mathcal{J}_{CH}(\omega_C + \omega_H) \right) +  \nonumber \\
		&& \frac{3}{4} d_{CD}^2 \left( 3 \mathcal{J}_{CH} (\omega_C)  + \mathcal{J}_{CH} (\omega_C - \omega_D)+ 6 \mathcal{J}_{CH}(\omega_C + \omega_D) \right),
\end{eqnarray}
\begin{eqnarray}
\mathrm{R}_2 =&& \frac{1}{18} \Delta \sigma_c^2 \omega_C^2 \left( 4 \mathcal{J}_C(0) + 3 \mathcal{J}_C(\omega_C) \right)\nonumber \\
		&&  \frac{1}{8} d_{CH}^2 \left(  4 \mathcal{J}_{CH}(0) + \mathcal{J}_{CH} (\omega_C - \omega_H)  + 6 \mathcal{J}_{CH}(\omega_H) + 6 \mathcal{J}_{CH}(\omega_C + \omega_H) + 3 \mathcal{J}_{CH}(\omega_C) \right) + \nonumber \\
		&& \frac{2}{3} d_{CD}^2 \left(  4 \mathcal{J}_{CH}(0) +  \mathcal{J}_{CH} (\omega_C - \omega_D)  +  6 \mathcal{J}_{CH}(\omega_D) + 6 \mathcal{J}_{CH}(\omega_C + \omega_D) + 3 \mathcal{J}_{CH}(\omega_C) \right),
\end{eqnarray}
\begin{equation}
\sigma^\mathrm{NOE} = \frac{1}{4} d_{CH}^2 \left(6 \mathcal{J}_{CH}(\omega_C + \omega_H) -\mathcal{J}_{CH}(\omega_C - \omega_H) \right),
\end{equation}
\begin{equation}
\eta_\mathrm{z} = \Delta \sigma_C d_{CH} \mathcal{J}_{CCH}(\omega_C),
\end{equation}
\begin{equation}
\eta_\mathrm{xy} = \frac{1}{6} \Delta \sigma_C d_{CH} \left(4 \mathcal{J}_{CCH}(0) + 3\mathcal{J}_{CCH}(\omega_C) \right),
\end{equation}
\end{subequations}
where the $d_{CX}$ is dipolar constant between the carbon-13 and nucleus X and equals $-\frac{\mu_0 \hbar \gamma_C \gamma_X}{4 \pi r_{CX}^3}$ with $\mu_0$ the permeability of free space, $\gamma_X$ the gyromagnetic ratio of nucleus X, $\hbar$ the Planck’s constant divided by $2\pi$ and $r_{CX}$ the internuclear distance, $\Delta \sigma_c$ is the carbon-13 CSA, $\omega_X = -\gamma_X B_0$ is the Larmor frequency for nucleus X at the magnetic field $B_0$ and the spectral density functions are defined as:
\begin{equation}
\mathcal{J}_I (\omega) = 2 \int_0^\infty C_I(t) e^{-i \omega t} dt,
\end{equation}
where the subscript \textit{I} refers to the considered correlation: to carbon-13 CSA auto-correlation when \textit{I}=\textit{C}, to DD interaction auto-correlation when \textit{I}=\textit{CX} and to carbon-13 CSA/CH DD cross-correlation when \textit{I}=\textit{CCH}. We assume that the methyl group adopts a perfect tetrahedral geometry with the carbon positioned at its center such that $\mathcal{J}_{CH} = \mathcal{J}_{CD}$.\\
In the case where the  CSA  tensors are  asymmetric and the strengths of the interactions are included in the correlation functions (as denoted by the superscript $(I)$) to account for their time-dependent fluctuations, these expressions are modified to:
\begin{subequations}
\label{eq:AllRatesMFCSAasym}
\begin{eqnarray}
\mathrm{R}_1 =&& \frac{1}{2} \left( \mathcal{J}^{(I)}_{\sigma_\parallel}(\omega_C)  + \mathcal{J}^{(I)}_{\sigma_\perp}(\omega_C) + 2 \mathcal{J}^{(I)}_{\sigma_\parallel, \sigma_\perp}(\omega_C) \right) + \nonumber \\
		&&  \frac{1}{4}  \left(\mathcal{J}^{(I)}_{CH} (\omega_C - \omega_H) +  3 \mathcal{J}^{(I)}_{CH} (\omega_C) + 6 \mathcal{J}^{(I)}_{CH}(\omega_C + \omega_H) \right) +  \nonumber \\
		&& \frac{3}{4} \left( 3 \mathcal{J}^{(I)}_{CD} (\omega_C)  + \mathcal{J}^{(I)}_{CD} (\omega_C - \omega_D)+ 6 \mathcal{J}^{(I)}_{CD}(\omega_C + \omega_D) \right),
\end{eqnarray}
\begin{eqnarray}
\mathrm{R}_2 =&& \frac{1}{12} \left( 8 \mathcal{J}^{(I)}_{\sigma_\parallel, \sigma_\perp}(0) + 6 \mathcal{J}^{(I)}_{\sigma_\parallel, \sigma_\perp}(\omega_C) + 4 \mathcal{J}^{(I)}_{\sigma_\parallel}(0) + 3 \mathcal{J}^{(I)}_{\sigma_\parallel}(\omega_C) + 4 \mathcal{J}^{(I)}_{\sigma_\perp}(0) + 3 \mathcal{J}^{(I)}_{\sigma_\perp}(\omega_C) \right) \nonumber \\
		&&  \frac{1}{8}  \left(  4 \mathcal{J}^{(I)}_{CH}(0) + \mathcal{J}^{(I)}_{CH} (\omega_C - \omega_H)  + 6 \mathcal{J}^{(I)}_{CH}(\omega_H) + 6 \mathcal{J}^{(I)}_{CH}(\omega_C + \omega_H) + 3 \mathcal{J}^{(I)}_{CH}(\omega_C) \right) + \nonumber \\
		&& \frac{2}{3}  \left(  4 \mathcal{J}^{(I)}_{CD}(0) +  \mathcal{J}^{(I)}_{CD} (\omega_C - \omega_D)  +  6 \mathcal{J}^{(I)}_{CD}(\omega_D) + 6 \mathcal{J}^{(I)}_{CD}(\omega_C + \omega_D) + 3 \mathcal{J}^{(I)}_{CD}(\omega_C) \right),
\end{eqnarray}
\begin{equation}
\sigma^\mathrm{NOE} = \frac{1}{4} \left(6 \mathcal{J}^{(I)}_{CH}(\omega_C + \omega_H) -\mathcal{J}^{(I)}_{CH}(\omega_C - \omega_H) \right),
\end{equation}
\begin{equation}
\eta_\mathrm{z} = \sqrt{\frac{3}{2}} \left(  \mathcal{J}^{(I)}_{\sigma_\parallel,CH}(\omega_C) +  \mathcal{J}^{(I)}_{\sigma_\perp,CH}(\omega_C) \right),
\end{equation}
\begin{equation}
\eta_\mathrm{xy} = \sqrt{\frac{1}{6}}  \left(4 \mathcal{J}^{(I)}_{\sigma_\parallel,CH}(0) + 3\mathcal{J}^{(I)}_{\sigma_\parallel,CH}(\omega_C) +  4 \mathcal{J}^{(I)}_{\sigma_\perp,CH}(0) + 3\mathcal{J}^{(I)}_{\sigma_\perp,CH}(\omega_C) \right),
\end{equation}
\end{subequations}
where $\mathcal{J}^{(I)}_{\sigma_i}$ refers to the spectral density function for the component $i$ of the carbon-13 CSA  tensor represented as the sum of two axially symmetric tensors,  $\mathcal{J}^{(I)}_{\sigma_\parallel, \sigma_\perp}$ is the spectral density function for  cross-correlation between the two CSA components and $\mathcal{J}^{(I)}_{\sigma_i, CH}$ is the spectral density function  for cross-correlation between the C-H DD interaction and the component $i$ of the carbon-13 CSA  tensor.

\section{Details on the computation of CSA tensors}
\begin{figure}[!ht]
	\begin{center}
		\includegraphics[width=0.9\textwidth]{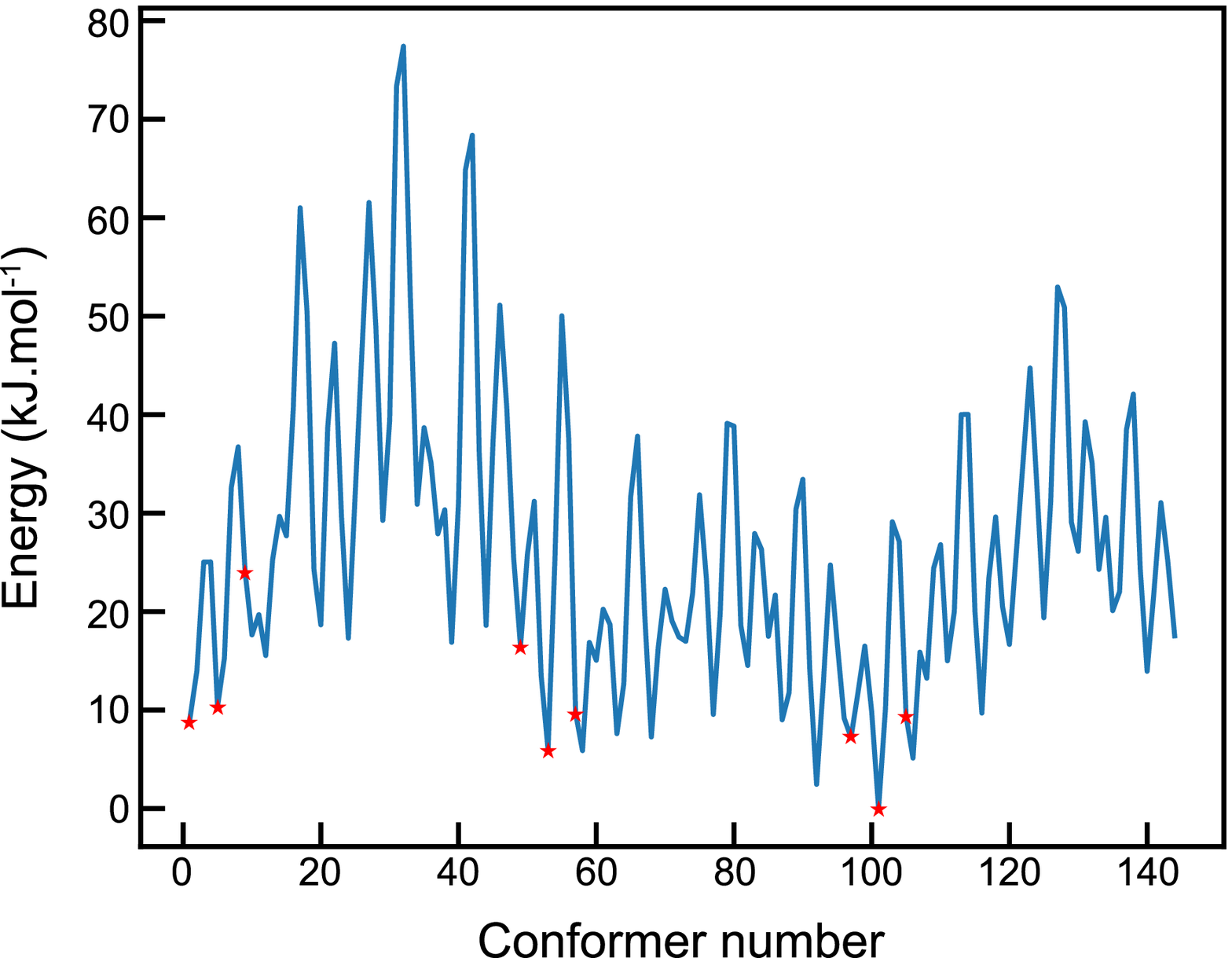}
	\end{center}
	\caption{Energies of the 144 conformations of an isoleucine amino acid in water and calculated using DFT. The conformation with the lowest energy (conformation 101) defines the reference point. The 9 selected rotamers are indicated with the red stars. Points are connected with a line for visual clarity.}
	\label{fig:DFTenergies}
\end{figure}

\paragraph{Geometry optimization of the rotamers} The conformation of an isoleucine side-chain is defined by the pair of dihedral angles ($\chi_1$, $\chi_2$), with $\chi_1$ being associated to the C$_\alpha$-N and C$_\beta$-C$_{\gamma1}$ bonds, and $\chi_2$ to the C$_\alpha$-C$_\beta$ and C$_{\gamma1}$-C$_{\delta1}$ bonds. Both of these dihedral angles can identify 3 staggered conformations, for values expected to be close to 60\,$^\circ$, 180\,$^\circ$  and 300\,$^\circ$, thus leading to 9 possible rotamer states. In order to obtain a structure for the 9 rotamers, an initial structure of isoleucine in zwitterionic form was optimized with DFT methods as implemented in Gaussian 09 Revision A.01 [6]. The B3LYP [7,8] hybrid functional and 6-311++G(2d,p) Pople basis set were chosen [9], with solvent effects due to water implicitly taken into account by means of the polarizable continuum model [10]. The local minimum produced by this geometry optimization resulted in $\chi_1$ and $\chi_2$ values of –63.05 and –64.85\,$^\circ$, respectively. Subsequently, this structure was utilized to perform a relaxed Potential Energy Surface (PES) scan at the same level of theory by varying independently the two dihedral angles in steps of 30\,$^\circ$ over 360\,$^\circ$, so as to generate 144 conformers. The 9 rotamers were thus identified as local minima of the cost function defined as:
\begin{equation}
f(\chi_1, \chi_2, r) = \sqrt{(\chi_{1,r} - \chi_1)^2 + (\chi_{2,r} - \chi_2)^2},
\end{equation}
were $\chi_{1,r}$ and $\chi_{2,r}$ are the theoretical $\chi_1$ and $\chi_2$ angles in rotamer $r$. The values of the $\chi_1$ and $\chi_2$ angles for the selected structures are reported in Table\,\ref{table:SelectedIleConf}. These rotamers also correspond to local minima of the PES, or are close to local minima, and fall within an energy window of \textit{ca.} 10.5\,kJ.mol$^{-1}$ (Fig.\,\ref{fig:DFTenergies}). The global minimum is characterized by $\chi_1 = 59.88$\,$^\circ$ and $\chi_2 = 173.85$\,$^\circ$.  \\

\paragraph{CSA calculation} The CSA tensors of the selected rotamers were calculated by means of the Gauge-Independent Atomic Orbital (GIAO) method [11,12] at the same level of theory as for the geometry optimizations. The CSA tensors were then diagonalized to extract the principal components and orientations of the 3 main axes in the molecular frame, and define the longitudinal and orthogonal components of the CSA interaction as we decompose the fully asymmetric CSA tensors into two axially symmetric interactions. In order to calculate the relaxation rates, the set of Euler angles for the orientation of the CSA components in the rotamer frame has to be determined. In order to do so, we first defined the jump frame with the z-axis pointing along the C$_\alpha$-C$_\beta$ bond, and the x-axis defined as:
\begin{equation}
\vec{x}_J = \frac{\overrightarrow{C_\alpha N} \wedge \vec{z}_J}{\lVert \overrightarrow{C_\alpha N} \rVert},
\end{equation}
where $\overrightarrow{C_\alpha N}$ is the vector pointing in the direction of the C$_\alpha$-N bond, and $\vec{x}_J$ and $\vec{z}_J$ are the normalized vectors defining the x- and z-axes of the jump frame. From this frame, the Euler angles $\varphi_{J,R}$ and $\theta_{J,R}$ defining the orientation of the rotamer frame in the jump frame can be calculated. In each rotamer frame, defined after applying transformations of the jump frame with the corresponding Euler angles, the orientation of the CSA components can be calculated. The amplitude of the CSA components and their respective set of Euler angles for orientation in the rotamer frames are given in Table\,\ref{table:CSAtensors}. \\

\section{Details on the MD simulation}
\paragraph{Simulation}
The topology for human ubiquitin (PDB: 1UBI) was processed using GROMACS 2018.4 tools and the AMBER 99SB*-ILDN force field [13,14]. The obtained protein topology file was then modified for accurate methyl group dynamics [15] using the script provided by Hoffmann \textit{et al.} (\url{https://github.com/fahoffmann/MethylRelax}). The system was solvated in a 6.5×6.5×6.5\,nm cubic box using the TIP4P-2005 [16] water model and 8 Na+/Cl- ion pairs, corresponding to a 50\,mM NaCl concentration and a total of 8787 water molecules. The small size of Ubiquitin ensures that the minimal distance between periodic images of the protein is at least twice as large as the interaction cutoff (1\,nm). Though hydrodynamic correlation effects might potentially affect rotational diffusion, no effects are expected regarding the internal dynamics of the protein, which represents the only aspect investigated here. The energy was minimized using gradient descent and short (100\,ps) equilibration runs in the NVT an NPT ensembles were used to bring the system at 300\,K and 1\,bar, with restraints on heavy atoms. \\
A 1\,$\mu$s production run was carried out using the leap-frog integrator with a 2\,fs timestep. Short-range van der Waals and Coulomb interactions were computed with a 1\,nm cutoff. Long-range electrostatics were treated using Particle Mesh Ewald. The protein and solvent were maintained at 300\,K using two separate V-rescale thermostats (coupling constant: 0.1\,ps). The pressure was fixed at 1\,bar using a Parrinello-Rahman barostat (coupling constant: 2.0\,ps). Analytical long-range dispersion corrections for energy and pressure were activated. The protein configurations were saved every 0.5\,ps in order to resolve fast methyl rotational jumps. \\

\paragraph{Calculation of the correlation functions}
The MD correlation functions were computed as follows. The C$_{\gamma 1}$-C$_{\delta 1}$ and C$_{\delta 1}$-H$_{\delta 1}$ bond vectors were extracted from the trajectories. The autocorrelation function $C_c(t)$ at times $0, \delta t, 2\delta t … N\delta t$ was then computed as:
\begin{equation}
C_a(k\delta t) = \frac{1}{N-k+1} \sum_{i=0}^{N-k}{P_2[\vec{u}(i\delta t) \cdot \vec{u}(i\delta t + k\delta t)]},
\end{equation}
where $\vec{u}(t)$ is the normalized bond vector at time $t$. We fixed N to 100,000 (50\,ns). The C$_{\delta1}$-H$_{\delta1}$ auto-correlation function used in this study corresponds to the average over the three protons of the methyl groups. The C$_{\gamma1}$-C$_{\delta1}$/C$_{\delta1}$-H$_{\delta1}$ cross-correlation function $C_c(t)$ was computed as:
\begin{equation}
C_c(k\delta t) = \frac{1}{N-k+1} \sum_{i=0}^{N-k}{P_2[\vec{u}(i\delta t) \cdot \vec{v}(i\delta t + k\delta t)]},
\end{equation}
where $\vec{u}(t)$ and $\vec{v}(t)$ refer to the normalized bond vectors orientated along the C$_{\gamma1}$-C$_{\delta1}$ and C$_{\delta1}$-H$_{\delta1}$ bond, respectively, at time $t$. \\
The initial drop of the MD correlation function was accounted for my discarding the first two points ($t \leq 1$\,ps) and normalizing the correlation functions to their values at $t = 1$\,ps. We used the scipy function \textit{curve\_fit} [17] to fit the MD correlation functions to a sum of exponential decays of the form:
 \begin{equation}
C_{fit} (t) =  A_0 + \sum_{i=1}^4 A_i e^{-t/\tau_i},
\end{equation}
where we impose $\sum_i A_i = 1$ in the auto-correlation case and $\sum_i A_i = C_c(0)$ in the cross-correlation case. The spectral density functions were then calculated as:
\begin{equation}
\mathcal{J}_{a/c} (\omega) = \frac{2}{5} \frac{A_0 \tau_c}{1 + (\omega \tau_c)^2} + \frac{2}{5} \sum_{i = 1}^4 \frac{A_i \tau_i'}{1 + (\omega \tau_i')^2},
\end{equation} 
where $\tau_i' = \tau_c \tau_i/(\tau_c + \tau_i)$. These functions can be used to compute relaxation rates. \\

\paragraph{Evaluation of jump rates}
Rotamer populations and jump rates were estimated from the MD trajectory by constructing a continuous time Markov State Model (MSM) for each residue. The rotameric states $(\chi_1, \chi_2)$ were discretized as shown in Fig.\,\ref{fig:Ramachandran} of the main text. To avoid artifacts from fast recrossings at the discretization boundaries, transitions between states were only accounted for if the side-chain was populating the new rotamer for at least 2.5\,ps (5 frames) after the transition. The trajectory in state space was analyzed in Python using the msmbuilder 3.8.0 [18] to obtain the (symmetrized) rate matrices and stationary probability distributions as well as the associated uncertainties corresponding to the asymptotic standard deviations. \\

\paragraph{Calculation of the conformational entropy}
Differential entropies in the $(\chi_1, \chi_2)$ space were computed from the trajectory using the Kozachenko and Leonenko k-nearest neighbor estimator ($k=3$) [19].

\newpage
\section{Additional figures}

\begin{figure*}[!ht]
\begin{center}
\includegraphics[width=0.9\textwidth]{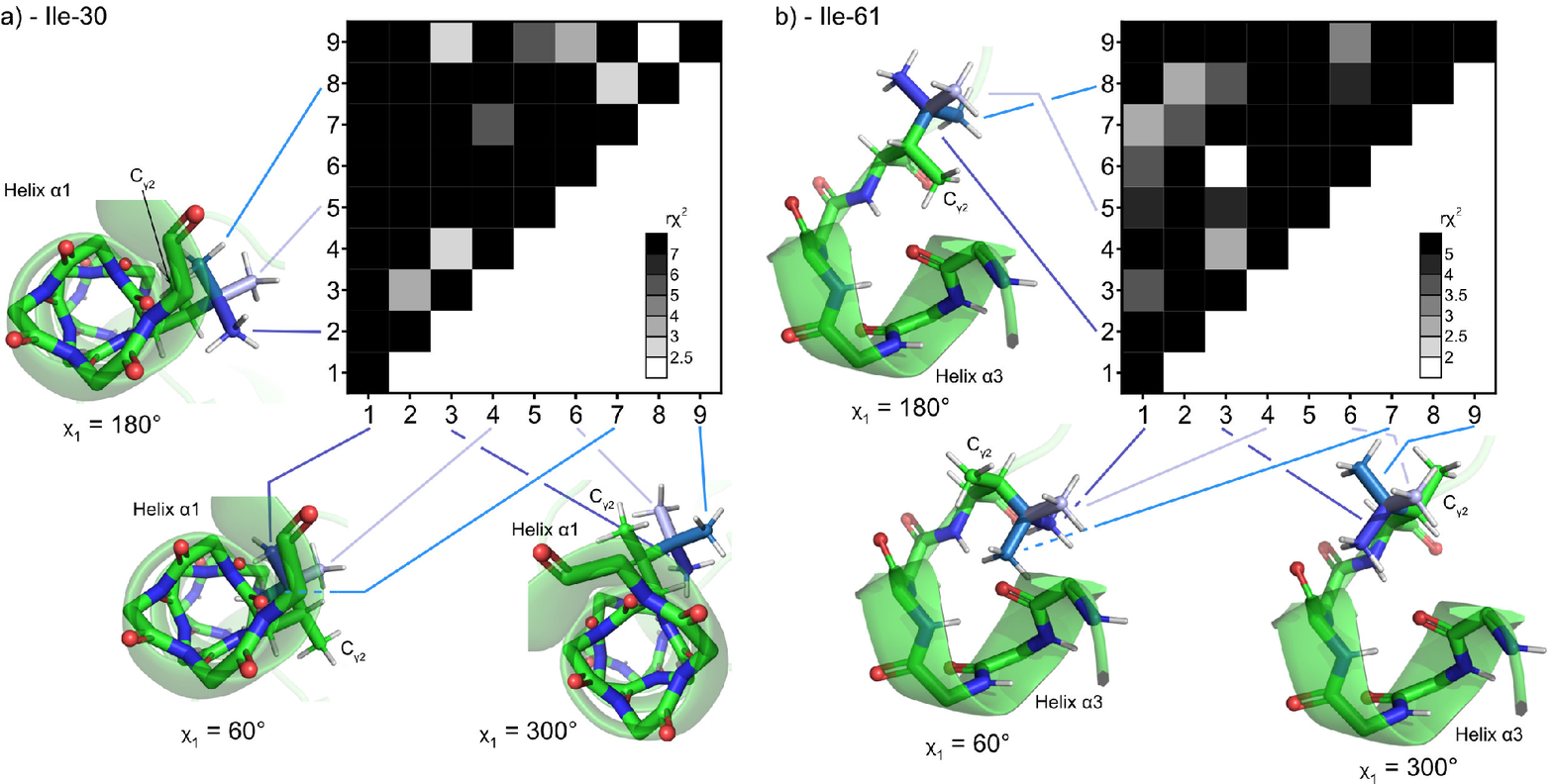}
\end{center}
\caption{\label{fig:61DifRot}
Analysis of the NMR relaxation rates with an explicit model of motions with different populated rotamer states for Ile-30 (a) and Ile-61 (b). The 2D plots show the variation of the reduced $\chi^2$ ($r\chi^2$) using a grey scale and as a function of the exchanging rotamer states. The rotamer numbers refer to Fig.\,\ref{fig:Ramachandran} of the main text. Note that the MD simulations indicates that rotamers 6 and 9 are populated for both residues. For both residues, the structure of the side-chain in the context of the protein is shown, with each sub-pannel showing an overlay for a given value of $\chi_1$ angle and the different orientations of the C$_{\delta1}$H$_3$ methyl groups distinguished by tints of blue. Rotamer states were generated by applying rotation matrices around the C$_\alpha$-C$_\beta$ bond (to generate different $\chi_1$ states) and C$_\beta$-C$_{\gamma1}$ bond (to generate different $\chi_2$ states) and using the model 1 of PDB 1D3Z structure as a starting point. Both isoleucines are in rotamer state number 6 in this model. Generating these fictional side-chain conformers highlights potential steric clashes with the protein backbone and populating such states would require major protein conformational changes.}
\end{figure*}

\clearpage
\newpage

\section{Additional tables}

\begin{table}[!ht]
\caption{\label{table:CSAtensors}
Amplitude ($\sigma$) and orientation (Euler angles $\varphi_{R,\sigma}$ and $\theta_{R,\sigma}$) of the longitudinal (denoted by the subscript $\parallel$) and orthogonal (denoted by the subscript $\perp$) components of the CSA tensors as calculated with DFT methods for the 9 rotamers. The isotropic chemical shifts ($\delta$) were referenced against TMS. The rotamers are numbered following Fig.\,\ref{fig:Ramachandran} of the main text.}
\begin{tabular}{|c|ccc|ccc|c|}
\hline
 & \multicolumn{3}{c|}{Longitudinal} & \multicolumn{3}{c|}{Orthogonal} &  \\
 r & $\sigma_\parallel$ (ppm) & $\varphi_{R,\sigma_\parallel}$ ($^\circ$) & $\theta_{R,\sigma_\parallel}$ ($^\circ$)  & $\sigma_\perp$ (ppm) & $\varphi_{R,\sigma_\perp}$ ($^\circ$) & $\theta_{R,\sigma_\perp}$ ($^\circ$) & $\delta (r)$ (ppm)\\
\hline
1 & 23.61 & 103.08 & 39.66 & 6.03 & 163.24 & 118.11 & 12.99  \\
2 & 26.32 & 253.69 & 47.17 & 6.74 & 157.87 & 73.93 & 13.77\\
3 & 22.69 & 226.33 & 151.17 & 6.63 & 166.75 & 93.24 & 15.09\\
4 & 25.81 & 176.84& 9.59& 7.11& 100.69& 95.61& 13.57\\
5 & 27.48 & 166.95 & 163.57 & 9.21 & 97.39 & 97.41 & 14.23\\
6 & 25.58 & 172.60 & 24.43 & 8.25 & 229.37 & 69.59 & 13.54\\
7 & 21.15 & 156.81 & 8.15 & 11.03 & 156.91 & 79.80 & 10.10\\
8 & 17.25 & 248.01 & 157.41 & 6.55 & 131.90 & 107.37 & 7.49\\
9 & 14.17 & 138.63 & 119.87 & 2.56 & 220.43 & 139.87 & 7.06\\
\hline
\end{tabular}
\end{table}

\begin{table*}[!ht]
\caption{\label{table:PopMD}
Fractional population of each rotamer according to the MD trajectory. We indicate rotamer states which are populated by less than 1\,\% with "-". The standard deviations were obtained from a Markov-state model analysis of the MD simulation.}
\begin{tabular}{|c|ccccc|}
\hline
 & \multicolumn{5}{c|}{Rotamer number} \\
Ile & 1 & 2 & 3 & 4 & 5 \\
\hline
3 & 0.05 $\pm$ 0.01 & 0.02 $\pm$ 0.02 & - & 0.79 $\pm$ 0.20 & 0.03 $\pm$ 0.04 \\
13 & 0.02 $\pm$ 0.0 & - & 0.12 $\pm$ 0.01 & 0.56 $\pm$ 0.04 & 0.01 $\pm$ 0.0  \\
23 & - & - & 0.03 $\pm$ 0.0 & - & - \\
30 & - & - & - & - & -  \\
36 & - & 0.05 $\pm$ 0.03 & 0.01  $\pm$ 0.0 & 0.05 $\pm$ 0.01 & 0.08 $\pm$ 0.05 \\
44 & - & - & 0.11 $\pm$ 0.0 & - & -  \\
61 & - & - & - & - & -  \\
\hline
\hline
Ile & 6 & 7 & 8 & 9&\\
\hline
3 &  - & - & - & 0.12 $\pm$ 0.16 & \\
13 &  0.26 $\pm$ 0.03 & - & - & 0.03 $\pm$ 0.0 & \\
23 &  0.06 $\pm$ 0.01 & - & - & 0.92 $\pm$ 0.01 &\\
30 & 0.92 $\pm$ 0.01 & - & - & 0.08 $\pm$ 0.01 & \\
36 & 0.59 $\pm$ 0.06 & 0.03 $\pm$ 0.01 & - & 0.19 $\pm$ 0.04 &\\
44 & 0.58 $\pm$ 0.02 & - & - & 0.31 $\pm$ 0.02 &\\
61 & 0.82 $\pm$ 0.02 & - & - & 0.18  $\pm$  0.02 &\\
\hline
\end{tabular}
\end{table*}

\begin{table}[!ht]
\caption{\label{table:Chi2All}
Reduced $\chi^2$ calculated for the different method of analysis presented here. The reduced $\chi^2$ is defined as $r\chi^2 = \frac{1}{(N -N_p)} \sum_i (R_i^{(c)} - R_i^{(m)})^2/\sigma_i^2$ where $R_i^{(c)}$ and $R_i^{(m)}$ are the calculated and measured rates, $\sigma_i$ is the experimental error, $N_p$ is the number of parameters in the model and the sum runs over the $N$ recorded rates. The column "MD - exp. decay" refers to the calculation of relaxation rates using a multi-exponential fit of the MD-derived correlation function (Fig.\,\ref{fig:MDrates_MF}) while "MD - explicit" refers to the calculation of relaxation rates when using an explicit model of motions with parameters obtained form the analysis of the MD trajectory (Fig.\,\ref{fig:MDrates_Explicit}). The number of parameter $N_p$ is set to 0 in the case where relaxation rates are directly calculated from the MD correlation rates.}
\begin{tabular}{|c|cccc|}
\hline
 residue & MF & MD - exp. decay & MD - explicit & explicit \\
\hline
3 & 3.67 & 190.6 & 11.9 & 3.76 \\
13 & 3.86 & 271.0 & 21.4 & 2.57 \\
23 & 5.09 & 353.9 & 90.2 & 2.45  \\
30 & 5.23 & 174.1 & 22.9 & 5.12 \\
36 & 10.86 & 347.8 & 49.5 & 1.27 \\
44 & 65.82 & 852.3 & 107.7 & 29.74 \\
61 & 9.70 & 244.2 & 104.5 & 3.84 \\
\hline
\end{tabular}
\end{table}

\begin{table*}[!ht]
\caption{\label{table:PopRelax}
Fractional population of each rotamer obtained from the analysis of NMR relaxation data using explicit models of motion. We indicate rotamer states which were not considered in the analysis with "-".}
\begin{tabular}{|c|ccc|ccc|ccc|ccc|ccc|}
\hline
Ile & p$_1$ & $+\sigma$ & $-\sigma$  & p$_2$ & $+\sigma$ & $-\sigma$ & p$_3$ & $+\sigma$ & $-\sigma$ & p$_4$ & $+\sigma$ & $-\sigma$ & p$_5$ & $+\sigma$ & $-\sigma$ \\
\hline
3 & 0.10 & 0.01 & 0.03 & 0.08 & 0.01 & 0.03 &  \multicolumn{3}{c|}{-} & 0.58 & 0.04 & 0.03 & 0.00 & 0.03 & 0.00  \\
13 & 0.08 & 0.01 & 0.02 &  \multicolumn{3}{c|}{-} & 0.14 & 0.01 & 0.01 & 0.28 & 0.07 & 0.05 & \multicolumn{3}{c|}{-}  \\
23 & \multicolumn{3}{c|}{-} & \multicolumn{3}{c|}{-} & 0.31 & 0.03 & 0.02 & \multicolumn{3}{c|}{-} & \multicolumn{3}{c|}{-} \\
30 & \multicolumn{3}{c|}{-} & \multicolumn{3}{c|}{-} & \multicolumn{3}{c|}{-} & \multicolumn{3}{c|}{-} & \multicolumn{3}{c|}{-} \\
36 & \multicolumn{3}{c|}{-} & 0.07 & 0.02 & 0.04 & \multicolumn{3}{c|}{-} & 0.09 & 0.04 & 0.05 & 0.22 & 0.09 & 0.09 \\
44 & \multicolumn{3}{c|}{-} & \multicolumn{3}{c|}{-} & 0.27 & 0.02 & 0.03 & \multicolumn{3}{c|}{-} & \multicolumn{3}{c|}{-}  \\
61 & \multicolumn{3}{c|}{-} & \multicolumn{3}{c|}{-} & \multicolumn{3}{c|}{-} & \multicolumn{3}{c|}{-} & \multicolumn{3}{c|}{-} \\
\hline 
\hline
Ile & p$_6$ & $+\sigma$ & $-\sigma$ & p$_7$ & $+\sigma$ & $-\sigma$ & p$_8$ & $+\sigma$ & $-\sigma$ & p$_9$ & $+\sigma$ & $-\sigma$ &&&\\
\hline
3 &  \multicolumn{3}{c|}{-} & \multicolumn{3}{c|}{-} & \multicolumn{3}{c|}{-} & 0.24 & 0.02 & 0.03 &&&\\
13 &  0.51 & 0.05 & 0.06 & \multicolumn{3}{c|}{-} & \multicolumn{3}{c|}{-} & \multicolumn{3}{c|}{-} &&& \\
23 & 0.08 & 0.00 & 0.01 & \multicolumn{3}{c|}{-} & \multicolumn{3}{c|}{-} & 0.61 & 0.01 & 0.02 &&&\\
30 & 0.90 & 0.00 & 0.00 & \multicolumn{3}{c|}{-} & \multicolumn{3}{c|}{-} & 0.10 & 0.00 & 0.00 &&&\\
36 & 0.45 & 0.08 & 0.07 & 0.05 & 0.01 & 0.01 & \multicolumn{3}{c|}{-} & 0.11 & 0.04 & 0.02 &&&\\
44 & 0.35 & 0.08 & 0.02 & \multicolumn{3}{c|}{-} & \multicolumn{3}{c|}{-} & 0.37 & 0.03 & 0.05 &&&\\
61 & 0.83 & 0.00 & 0.11 & \multicolumn{3}{c|}{-} & \multicolumn{3}{c|}{-} & 0.17 & 0.11 & 0.00 &&&\\
\hline
\end{tabular}
\end{table*}

\begin{table}[!ht]
	\caption{Configuration entropies ($S_c$ and order parameters for rotamer jumps ($\mathcal{S}_J^2$) determined from the analysis of the NMR relaxation data. The uncertainties correspond to the width of the distributions after calculating the entropy and order parameter for each steps of the MCMC trajectory.}%
	\begin{center}
		{\def\arraystretch{1.5}
		\begin{tabular}{|c|ccc|ccc|}
			\hline
		 	Ile  & \multicolumn{3}{c|}{$S_c/k_B$} & \multicolumn{3}{c|}{$\mathcal{S}_J^2$}  \\%
			\hline
			 & mean & $+\sigma$ & $-\sigma$ & mean & $+\sigma$ & $-\sigma$   \\ 
			\hline
			3 & 1.12 & 0.03 & 0.05 & 0.25 &  0.06 & 0.02 \\
			13 & 1.17 & 0.02 & 0.03 & 0.46 & 0.01 & 0.01 \\
			23 & 0.87 & 0.01 & 0.01 & 0.53 & 0.01 & 0.00 \\
			30 & 0.33 & 0.01 & 0.01 & 0.73 & 0.01 & 0.01 \\
			36 & 1.45 & 0.08 & 0.08 & 0.21 & 0.10 & 0.07 \\
			44 & 1.09 & 0.01 & 0.02 & 0.19 & 0.01 & 0.02 \\
			61 & 0.45 & 0.14 & 0.01 & 0.58 & 0.01 & 0.18 \\
			\hline
		\end{tabular}
		}
	\end{center}
	\label{table:Entropie}
\end{table}

\begin{table}[!ht]
	\caption{Values of $\chi_1$ and $\chi_2$ angles obtained after DFT optimization of isoleucine structures, and defining the 9 rotamers, and Euler angles $\varphi_{J,R}$ and $\theta_{J,R}$ defining the orientation of the rotamer frame in the jump frame. }%
	\begin{center}
		{\def\arraystretch{1.5}
		\begin{tabular}{|c|cc|cc|cc|}
			\hline
		 	  & \multicolumn{2}{c|}{$\chi_1$ ($^\circ$)} & \multicolumn{2}{c|}{$\chi_2$ ($^\circ$)} & &  \\%
			\hline
			Rotamer & Theoretical & DFT & Theoretical & DFT & $\varphi_{J,R}$ ($^\circ$) & $\theta_{J,R}$ ($^\circ$)  \\ 
			\hline
			1 & 60 & 55.58 & 60 & 56.70 & 105.33 & 103.74 \\
			2 & 180 & 178.08 & 60 & 53.63 & 228.21 & 108.71 \\
			3 & 300 & 291.44 & 60 & 59.22 & 199.06 & 101.55 \\
			4 & 60 & 59.88 & 180 & 173.85 & 135.10 & 5.80 \\
			5 & 180 & 180.25 & 180 & 172.19 & 249.12 & 7.65 \\
			6 & 300 & 295.36 & 180 & 177.00 & 211.35 & 4.97 \\
			7 & 60 & 63.59 & 300 & 289.81 & 172.20 & 92.98 \\
			8 & 180 & 184.40 & 300 & 287.02 & 127.09 & 91.56 \\
			9 & 300 & 296.95 & 300 & 295.15 & 241.33 & 100.43 \\
			\hline
		\end{tabular}
		}
	\end{center}
	\label{table:SelectedIleConf}
\end{table}

\begin{table*}[ht]
\caption{\label{table:ParamDef}
Definitions of the parameters used in Eq.\,\ref{eq:AllCorrFunc}.}
\begin{tabular}{|c|c|}
\hline
 Label & Definition \\
\hline
$d_{CX }$ & dipolar coefficient for the $^{13}$C and X=$^1$H,$^2$H: $d_{CX} = -\frac{\mu_0}{4\pi}\frac{\hbar \gamma_C \gamma_X}{r_{CX}^3}$ \\
$\mu_0$ & permeability of free space \\
$\gamma_X$ & gyromagnetic ratio of nucleus X \\
$\hbar$ & Planck's constant divided by $2\pi$ \\
$r_{CX}$ & internuclear distance between $^{13}$C and X=$^1$H,$^2$H \\
$D_\parallel$ & longitudinal component of the diffusion tensor \\
$D_\perp$ & orthogonal component of the diffusion tensor \\
$d_{mn} $ & reduced Wigner matrix with index $m$ and $n$ \\
$p_\gamma$ & fractional population of rotamer $\gamma$, $\sum_\gamma p_\gamma =1$ \\
$\lambda_n$ & eigenvalue of the symmetrized exchange matrix $\tilde{\mathcal{R}}$ \\
$\tilde{X}_\gamma^{n}$ & $\gamma^\mathrm{th}$ value of the eigenvector associated to $\lambda_n$ \\
$\tilde{\mathcal{R}}$ & $\tilde{R}_{\alpha\beta} = \sqrt{ k_{\alpha \beta} k_{\beta \alpha}}$ for $\alpha \neq \beta$ and $\tilde{R}_{\alpha \alpha} = - \sum_{\beta \neq \alpha} k_{\beta \alpha}$ \\
$k_{\alpha \beta}$ & jump rate from rotamer $\beta$ to $\alpha$ \\
$N$ & number of rotamer states \\
$\Omega_{D,R_\gamma}$ & Euler angle orienting the rotamer frame $\gamma$ in the diffusion frame \\
$\Omega_{D,\sigma_i^{\gamma}}$ & Euler angle orienting the component $i$ of the CSA tensor in the diffusion frame \\
$D_m$ & diffusion coefficient for methyl rotation \\
$\beta_m$ & angle-opening of the methyl group (109.47\,$^\circ$) \\
$\Delta \sigma_i^{\gamma}$ & value of the component $i$ of the CSA tensor in rotamer $\gamma$ \\
$\omega_C$ & Larmor frequency for $^{13}$C \\
\hline
\end{tabular}
\end{table*}

\begin{table*}[!ht]
	\caption{Methyl-rotation correlation times calculated as $\tau_m = \frac{1}{1-\mathcal{S}_m^2} \sum_{a\neq0} \frac{d_{a,0}(\beta_m)^2}{a^2 D_m}$ with $\mathcal{S}_m^2=\mathcal{P}_2(\cos \beta_m)^2$ the squared methyl-rotation order parameter, $\mathcal{P}(x)=(3x^2-1)/2$ the second-order Legendre polynomial, $\beta_m$ the angle-opening of the methyl group (109.47$^\circ$) and $D_m$ the methyl-rotation diffusion coefficient obtained from analysis of the NMR relaxation rates. We report two values for Ile-23, the first one being associated to the rotation in rotamers 3 and 6, the second one to the rotation in rotamer 9.}%
	\begin{center}
		{\def\arraystretch{1.5}
		\begin{tabular}{|c|ccccccc|}
			\hline
		 	residue  & 3 & 13 & 23 & 30 & 36 & 44 & 61 \\%
			\hline
			$\tau_m$ (ps) & 10.1 & 12.7 & 9.2 / 44.6 & 8.3 & 8.4 & 5.7 & 15.3 \\
			\hline
		\end{tabular}
		}
	\end{center}
	\label{table:MetRotCorrTimes}
\end{table*}

\clearpage

\section{Jump matrices obtained from the analysis of NMR relaxation data}

We  report here the non-symmetrized jump matrices with rates obtained from the MCMC  analysis and using approximation following the  analysis  of  the MD  simulation. We write the jump rates as $\tilde{k}^{+\sigma}_{-\sigma}$ with $\tilde{k}$ the mean value and $+\sigma$ and $-\sigma$ the widths of the MCMC distributions.
\begin{subequations}
\label{eq:AllRatesMF}
\begin{equation}
\mathcal{R}_3  =
\begin{pmatrix}
-6.91^{3.42}_{2.13}\times10^{10} & 0 & 1.17^{0.39}_{0.57} \times10^{10} & 0 &  0 \\
0 & -2.60^{2.22}_{61.3}\times10^{8} & 0 &  6.08^{13.8}_{5.76} \times 10^9 & 1.10^{4.30}_{0.76}\times10^{7}\\
6.91^{2.13}_{3.42} \times 10^{10} & 0 & -1.17^{0.57}_{0.39} \times 10^{10} & 0 & 0 \\
0 & 1.04^{6.02}_{0.91} \times 10^8 & 0 & -6.16^{5.68}_{13.8} \times 10^9 & 8.98^{136}_{8.27} \times 10^5 \\
0 & 3.34^{11.9}_{2.02} \times 10^7 & 0 & 3.44^{11.1}_{2.08} \times 10^7 & -1.55^{1.07}_{5.71} \times 10^7
\end{pmatrix}
\end{equation}
\begin{equation}
\mathcal{R}_{13}  =
\begin{pmatrix}
-7.77^{2.53}_{2.62} \times 10^8 & 0 & 2.04^{1.03}_{0.70} \times 10^8 & 0 \\
0 & -1.55^{0.80}_{1.71} \times 10^{10} & 0 & 4.23^{5.17}_{1.82} \times 10^9 \\
7.77^{2.62}_{2.53} \times 10^8 & 0 & -2.04^{0.70}_{1.03} \times 10^8 & 0 \\
0 & 1.55^{1.71}_{0.80} \times 10^{10} & 0 & 4.23^{1.82}_{5.17} \times 10^9
\end{pmatrix}
\end{equation}
\begin{equation}
\mathcal{R}_{23}  =
\begin{pmatrix}
-1.45^{0.65}_{1.89} \times 10^{10} & 8.48^{16.8}_{3.45} \times 10^8 & 7.42^{9.27}_{3.42} \times 10^9 \\
2.33^{3.09}_{0.88} \times 10^8 & -6.59^{0.92}_{2.79} \times 10^9 & 8.14^{2.43}_{1.47} \times 10^8 \\
1.42^{1.90}_{0.65} \times 10^{10} & 5.77^{2.12}_{0.97} \times 10^9 & -8.24^{3.32}_{9.06} \times 10^9
\end{pmatrix}
\end{equation}
\begin{equation}
\mathcal{R}_{30}  =
\begin{pmatrix}
-1.08^{0.24}_{0.61} \times 10^9 & 9.66^{5.48}_{2.23} \times 10^{9} \\
1.08^{0.61}_{0.24} \times 10^9 & -9.66^{2.23}_{5.48} \times 10^{9}
\end{pmatrix}
\end{equation}
\begin{equation}
\mathcal{R}_{36}  =
\begin{pmatrix}
-1.16^{0.93}_{0.94} \times 10^{10} & 0 & 4.07^{5.24}_{3.47} \times 10^9 & 0 & 0 & 0 \\
0 & -1.76^{1.24}_{14.9} \times 10^9 & 0 & 3.22^{19.0}_{2.80} \times 10^8 & 4.10^{3.82}_{1.69} \times 10^8 & 0 \\
1.16^{0.94}_{0.93} \times 10^9 & 0 & -4.57^{2.68}_{10.0} \times 10^9 & 7.00^{18.5}_{5.89} \times 10^8 & 0 & 0 \\
0 & 1.62^{14.7}_{1.44} \times 10^9 & 1.03^{6.55}_{0.75} \times 10^8 & -2.41^{1.29}_{4.15} \times 10^9 & 0 & 2.79^{1.42}_{0.87} \times 10^9 \\
0 & 2.12^{5.13}_{1.13} \times 10^8 & 0 & 0 & -4.10^{1.69}_{3.82} \times 10^8 & 0 \\
0 & 0 & 0 & 6.58^{5.14}_{2.03} \times 10^8 & 0 & -2.79^{0.87}_{1.42} \times 10^9
\end{pmatrix}
\end{equation}
\begin{equation}
\mathcal{R}_{44}  =
\begin{pmatrix}
-8.66^{2.53}_{1.20} \times 10^9 & 1.27^{21.5}_{0.80} \times 10^8 & 6.10^{1.40}_{1.89} \times 10^9 \\
1.82^{2.97}_{1.15} \times 10^8 & -1.42^{0.16}_{0.14} \times 10^9 & 1.21^{0.24}_{0.18} \times 10^9 \\
8.45^{1.21}_{2.78} \times 10^9 & 1.25^{0.20}_{0.24} \times 10^9 & -7.28^{1.80}_{1.44} \times 10^9
\end{pmatrix}
\end{equation}
\begin{equation}
\mathcal{R}_{61}  =
\begin{pmatrix}
-1.36^{0.17}_{0.29} \times 10^9 & 6.39^{1.03}_{0.91} \times 10^9 \\
1.36^{0.29}_{0.17} \times 10^9 & -6.39^{0.91}_{1.03} \times 10^9
\end{pmatrix}
\end{equation}
\end{subequations}

\section{Additional bibliography}

[1] S. F. Cousin, P. Kade\v{r}\'{a}vek, N. Bolik-Coulon, Y. Gu, C. Charlier, L. Carlier, L. Bruschweiler-Li, T. Marquardsen, J.-M. Tyburn, R. Br{\"{u}}schweiler, F. Ferrage, J.\,Am.\,Chem.\,Soc. \textbf{140}, 13456 (2018) \\

\noindent[2] D. Fushman, R. Xu, D. Cowburn, Biochemistry \textbf{38}, 10225 (1999) \\

\noindent[3] C. Charlier, S. N. Khan, T. Marquardsen, P. Pelupessy, V. Reiss, D. Sakellariou, G. Bodenhausen, F. Engelke, F. Ferrage, J.\,Am.\,Chem.\,Soc. \textbf{135}, 18665 (2013) \\

\noindent[4] N. Bolik-Coulon, F. Ferrage, J.\,Chem.\,Phys (2022), doi:10.1063/5.0095910 \\

\noindent[5] R. Zwanzig, Proc.\,Natl.\,Acad.\,Sci.\,USA \textbf{85}, 2029 (1988) \\

\noindent[6] M. J. Frisch, G. W. Trucks, H. B. Schlegel, G. E. Scuseria, M. A. Robb, J. R. Cheese- man, G. Scalmani, V. Barone, B. Mennucci, G. A. Petersson, H. Nakatsuji, M. Caricato, X. Li, H. P. Hratchian, A. F. Izmaylov, J. Bloino, G. Zheng, J. L. Sonnenberg, M. Hada, M. Ehara, K. Toyota, R. Fukuda, J. Hasegawa, M. 
Ishida, T. Nakajima, Y. Honda, O. Ki- tao, H. Nakai, T. Vreven, J. A. Montgomery Jr., J. E. Peralta, F. Ogliaro, M. Bearpark, J. J. Heyd, E. Brothers, K. N. Kudin, V. N. Staroverov, R. Kobayashi, J. Normand, K. Raghavachari, A. Rendell, J. C. Burant, S. S. Iyengar, J. Tomasi, M. Cossi, N. Rega, J. M. Millam, M. Klene, J. E. 
Knox, J. B. Cross, V. Bakken, C. Adamo, J. Jaramillo, R. Gomperts, R. E. Stratmann, O. Yazyev, A. J. Austin, R. Cammi, C. Pomelli, J. W. Ochterski, R. L. 
Martin, K. Morokuma, V. G. Zakrzewski, G. A. Voth, P. Salvador, J. J. Dannenberg, S. Dapprich, A. D. Daniels, {\"{O}}. Farkas, J. B. Foresman, J. V. Ortiz, J. 
Cioslowski, D. J. Fox, Gaussian 09 Revision A.01 (2009), gaussian  Inc. Wallingford CT. \\

\noindent[7] C. Lee, W. Yang, R. G. Parr,  Phys.\ Rev.\ B \textbf{37}, 785  (1988) \\

\noindent[8] A. D. Becke, J.\ Chem.\ Phys. \textbf{98}, 5648 (1993) \\

\noindent[9] R. Ditchfield, W. J. Hehre, J. A. Pople, J.\ Chem.\ Phys. \textbf{54}, 724 (1971) \\

\noindent[10] J. Tomasi, B. Mennucci, R. Cammi, Chem.\ Rev. \textbf{105}, 2999 (2005) \\

\noindent[11] D. Zeroka, H. F. Hameka, J.\ Chem.\ Phys. \textbf{43}, 300 (1966) \\

\noindent[12] R. Ditchfield, J.\ Chem.\ Phys. \textbf{56}, 5688 (1972)  \\

\noindent[13] R. B. Best, G. Hummer, J.\,Phys.\,Chem.\,B \textbf{113}, 9004 (2009) \\

\noindent[14] K. Lindorff-Larsen, S. Piana, K. Palmo,. P. Maragakis, J. L. Klepeis, R. O. Dror, D. E. Shaw, Proteins \textbf{78}, 1950 (2010) \\

\noindent[15] F. Hoffmann, F. A. A. Mulder, L. V. Sch{\"{a}}fer, J.\,Phys.\,Chem.\,B \textbf{122}, 5038 (2018) \\

\noindent[16] L. J. F. Abascal, C. Vega, J.\,Chem.\,Phys \textbf{123}, 234505 (2005) \\

\noindent[17] P. Virtanen, R. Gommers, T. E. Oliphant, M. Haberland, T. Reddy, D. Cournapeau, E. Burovski, P. Peterson, W. Weckesser, J. Bright, S. J. van der Walt, M. Brett, J. Wilson, K. Jarrod Millman, N. Mayorov, A. R. J. Nelson, E. Jones, R. Kern, E. Larson, C. J. Carey, I. Polat, Y. Feng, E. W. Moore, J. VanderPlas, D. Laxalde, J. Perktold, R. Cimrman, I. Henriksen, E. A. Quintero, C. R. Harris, A. M. Archibald, A. H. Ribeiro, F. Pedregosa, P. van Mulbregt, Scipy 1.0 Contributors, Nat\,Methods \textbf{17}, 261 (2020) \\

\noindent[18] K. A. Beauchamp, G. R. Bowman, T. J. Lane, L. Maibaum, I. S. Haque, V. S. Pande, J.\,Chem.\,Theory\,Comput. \textbf{7}, 3412 (2011) \\

\noindent[19] H. Singh, N. Misra, A. Fedorowicz, E. Demchuk, Am.\,J.\,Math.\,Manag.\,Sci. \textbf{23}, 301 (2003) \\

\noindent[20] N. Bolik-Coulon, P. Kade\v{r}\'{a}vek, P. Pelupessy, J.-N. Dumez, F. Ferrage, S. F. Cousin, J.\,Magn.\,Reson. \textbf{313}, 106718 (2020)

\end{document}